\documentclass[10pt,letterpaper]{article}
\usepackage[top=0.85in,left=2.75in,footskip=0.75in]{geometry}

\usepackage{changepage}

\usepackage[utf8]{inputenc}

\usepackage{textcomp,marvosym}

\usepackage{fixltx2e}

\usepackage{amsmath,amssymb}

\usepackage{cite}

\usepackage{nameref,hyperref}

\usepackage{microtype}
\DisableLigatures[f]{encoding = *, family = * }

\usepackage{rotating}

\usepackage{subfig}

\usepackage{color}

\raggedright
\makeatletter
\renewcommand{\@biblabel}[1]{\quad#1.}
\makeatother

\date{}

\usepackage{lastpage,fancyhdr,graphicx}
\usepackage{epstopdf}
\fancyheadoffset[L]{2.25in}
\fancyfootoffset[L]{2.25in}

\newcommand{\argmax}{\operatornamewithlimits{argmax}}

\begin{document}
\vspace*{0.35in}

\begin{flushleft}
{\Large
\textbf\newline{Information processing features can detect behavioral regimes of dynamical systems}
}
\newline
\\
Rick Quax\textsuperscript{1,*},
Gregor Chliamovitch\textsuperscript{2},
Alexandre Dupuis\textsuperscript{2},
Jean-Luc Falcone\textsuperscript{2},
Bastien Chopard\textsuperscript{2},
Alfons G. Hoekstra\textsuperscript{1,3},
Peter M.A. Sloot\textsuperscript{1,3,4}
\\
\bigskip
\bf{1} Computational Science Lab, University of Amsterdam, The Netherlands
\\
\bf{2} Department of Computer Science, University of Geneva, Switzerland
\\
\bf{3} ITMO University, Saint Petersburg, Russia
\\
\bf{4} Complexity Institute, Nanyang Technological University, Singapore
\\
\bigskip

%
%





* r.quax@uva.nl

\end{flushleft}
\section*{Abstract}

In dynamical systems, local interactions between dynamical units generate correlations which are stored and transmitted throughout the system, generating the macroscopic behavior. However a framework to quantify and study this at the microscopic scale is missing. Here we propose an ‘information processing’ framework based on Shannon mutual information quantities between the initial and future states. We apply it to the 256 elementary cellular automata (ECA), which are the simplest possible dynamical systems exhibiting behaviors ranging from simple to complex.
Our main finding for ECA is that only a few features are needed for full predictability and that the `information integration' (synergy) feature is always most predictive. Finally we apply the formalism to foreign exchange (FX) and interest-rate swap (IRS) time series data and find that the 2008 financial crisis marks a sudden and sustained regime shift (FX and EUR IRS) resembling tipping point behavior. The USD IRS market exhibits instead a slow and steady progression which appears consistent with the hypothesis that this market is (part of) the driving force behind the crisis. Our work suggests that the proposed framework is a promising way of predicting emergent complex systemic behaviors in terms of the local information processing of units.

\section*{Introduction}

Emergent, complex behavior can arise from the interactions among (simple) dynamical units. An example is the brain whose complex behavior as a whole cannot be explained by the dynamics of a single neuron. In such a system, each dynamical unit receives input from other (upstream) units and then decides its next state, reflecting these correlated interactions. This new state is then used by (downstream) neighboring units to decide their new states, and so on, eventually generating a macroscopic behavior with systemic correlations. A quantitative framework is missing to fully understand how local interactions lead to complex emergent behaviors, or to predict whether a given system of local interactions will eventually generate complex systemic behavior or not.

Our hypothesis is that Shannon's information theory \cite{cover_elements_1991} can be used to construct such a framework. In this viewpoint, a unit's new state reflects its past interactions in the sense that it stores mutual information about the past states of upstream neighboring units. In the next time instant a downstream neighboring unit interacts with this state, implicitly transferring this information and integrating it together with other information into its new state, and so on. In effect, each interaction among dynamical units is interpreted as a Shannon communication channel and we aim to trace the onward transmission and integration of information through this network of `communication channels'.

In this paper we characterize the information in a single unit's state at time $t$ by enumerating its mutual information quantities with all possible sets of initial unit states ($t=0$). Then we quantify `information processing' as the progression of a unit's vector of information quantities over time (see Methods). We first test whether this notion of information processing could be used to predict if local interactions will generate complex emergent behavior in the theoretical framework of elementary cellular automata (ECA). Next we also test if information processing could be used to detect a difference of systemic behavior in real financial time series data, namely the regimes before and after the 2008 crisis.

The study of `information processing' of dynamical systems is a young and growing research topic. As illustrative examples, Lizier et al. propose a framework to formulate dynamical systems in terms of distributed ‘local’ computation: information storage, transfer, and modification \cite{lizier_information_2010} defined by individual terms of the Shannon mutual information sum (see Eq.~\ref{eq:mi}). For cellular automata they provide evidence for the long-held conjecture that so-called particle collisions are the primary mechanism for locally modifying information, and for a networked variant they show that a phase transition is characterized by the shifting balance of locally information storage over transfer \cite{lizier_information_2008}. A crucial difference with our work is that we operate in the ensemble setting, whereas Lizier et al. study a single realization of a dynamical system. Williams et al. trace how task-relevant information flows through a minimally cognitive agent’s neurons and environment to ultimately be combined into a categorization decision \cite{beer_information_2014} or sensorimotor behavior \cite{izquierdo_information_2015}. Studying how local interactions lead to multi-scale systemic behavior is also a domain which benefits from information-theoretic approaches, such as by Bar-Yam et al. \cite{bar-yam_computationally_2013,allen_information-theoretic_2014}, Quax et al. \cite{quax_diminishing_2013,quax_information_2013-1}, and Lindgren \cite{kristian_lindgren_information-theoretic_2015}. Finally, extending information theory itself to deal with complexity, multiple authors are concerned with decomposing a single information quantity into multiple constituents, such as synergistic information, including James et al.~\cite{james_anatomy_2011}, Williams and Beer~\cite{williams_nonnegative_2010}, Olbrich et al.~\cite{olbrich_information_2015}, Quax et al. \cite{quax_synergy}, Chliamovitch et al.~\cite{chliamovitch2014assessing}, and Griffith et al.~\cite{griffith_intersection_2014,griffith_synergy_entropy_2015}.


\section*{Methods}

\subsection*{Model of dynamical systems}
In general we consider discrete-time, discrete-state Markov dynamics. Let $X^{t} \equiv (X^{t}_{1}, X^{t}_{2}, \ldots, X^{t}_{N})$ denote the stochastic variable of the system state defined as the sequence of $N$ unit states at time $t$. Each unit chooses its new state locally according to the conditional probability distribution $\Pr_{i}(X^{t+1}_{i} \mid X^{t})$, encoding the microscopic system mechanics where $i$ identifies the unit. The state space of each unit is equal and denoted by the set $\Sigma$. We assume that the number of units, the system mechanics, and the state space remain unchanged over time. Finally we assume that all unit states are initialized identically and independently (i.i.d.), i.e., $\Pr(X^{0}) = \prod_{i=1}^{N} \Pr(X^{0}_i)$. The latter ensures that all correlations in future system states are generated by the interacting units and not an artifact of the initial conditions.

\paragraph{Elementary Cellular Automata.} Specifically we focus on the set of 256 elementary cellular automata (ECA) which are the simplest discrete spatio-temporal dynamical systems possible~\cite{wolfram_new_2002}. Each unit has two possible states and chooses its next state deterministically using the same transition rule as all other cells. The next state of a cell deterministically depends only on its own previous state and that of its two nearest neighbors, forming a line network of interactions. That is, 

\begin{equation}
{\Pr}_{r}(X^{t+1}_{i} \mid X^{t}) = {\Pr}_{r}(X^{t+1}_{i} \mid X^{t}_{i-1}, X^{t}_{i}, X^{t}_{i+1}).
\label{eq:condprob}
\end{equation}
There are 256 possible transition rules and they are numbered 0 through 255, denoted $r \in 0..255$. As initial state we take the fully random state so that no correlations exist already at $t=0$, i.e., $\Pr_r(X^{t=0}_{i})=1/2$ for all $r$ and all $i$. The evolution of each cellular automaton is fully deterministic for a given rule, implying that the conditional probabilities in Eq.~\ref{eq:condprob} can only be either $0$ or $1$. This is not a necessary condition for our framework.


\subsection*{Quantifying the information processing in a dynamical model}

\paragraph{Basics of information theory.} We interpret each interaction $\Pr_{i}(X^{t+1}_{i} \mid X^{t})$ as a set of Shannon communication channels, where each channel communicates information from a subset of $X^{t}$ to $X^{t+1}_{i}$. In general, a communication channel $A \rightarrow B$ between two stochastic variables is defined by the one-way interaction $\Pr(B \mid A)$ and is characterized by the amount of information about the state $A$ which transfers to the state $B$ due to this interaction. The average amount of information stored in the sender's state $A$ is determined by its marginal probability distribution $\Pr(A)$, which is known as its Shannon entropy:

\begin{equation}
H(A) =  - \sum\limits_a {\Pr (A = a)\log_2 \Pr (A = a)} .
\end{equation}
After a perfect, noiseless transmission, the information at the receiver $B$ will share exactly $H(A)$ bits with the information stored at the sender $A$. After a failed transmission the receiver shares zero information with the sender, and for noisy transmission their mutual information is somewhere in between. This is quantified by the so-called mutual information:
	  
\begin{equation}
I(A:B) = H(A) - H(A \mid B) = \sum\limits_{a,b} {\Pr (A = a,B = b)\log_2 \frac{{\Pr (A = a,B = b)}}{{\Pr (A = a)\Pr (B = b)}}}.
\label{eq:mi}
\end{equation}
The conditional variant $H(A \mid B)$ obeys the chain rule $H(A,B) = H(B) + H(A \mid B)$ and is written explicitly as

\[H(A \mid B) =  - \sum\limits_{b } {\Pr (B = b)\sum\limits_{a} {\Pr (A = a \mid B = b){{\log }_2}\Pr (A = a \mid B = b)} } .\]
%
This denotes the remaining entropy (uncertainty) of $A$ given that the value for $B$ is observed. For intuition it is easily verified that the case of statistical independence, i.e., $P(B{\mid}A)=P(B)$, leads to $H(A{\mid}B)=H(A)$ which makes $I(A:B)=0$, meaning that $B$ contains zero information about $A$. At the other extreme, $B=A$ would make $H(A{\mid}B)=0$ so that $I(A:B)=H(A)$, meaning that $B$ contains the maximal amount of information needed to determine a unique value of $A$.

\paragraph{Characterizing the information stored in a unit's state.} First we characterize the information stored in a unit's state at time step $t$, denoted $\vec{i}(X^{t}_{i})$, as the ordered sequence of mutual information quantities with all possible sets of unit states at time $t=0$, i.e.,

\begin{equation}
\vec{i}(X^{t}_{i}) \equiv (I(X^{t}_{i}:s) : s \in 2^{X^{0}}).
\end{equation}
Here $2^{X^{0}}$ is the power set notation for all subsets of initial cell state variables. We will refer to $\vec{i}(X^{t}_{i})$ as the sequence of information features of unit $i$ at time $t$.

In particular we highlight the following three types of information features. The `memory' of unit $i$ at time $t$ is defined as the features $I(X^{t}_{i}:X^{0}_{i}) \in \vec{i}(X^{t}_{i})$, i.e., the amount of information that the unit retains about its own initial state. The `transfer' of information is defined as non-local mutual information such as $I(X^{t}_{i}:X^{0}_{j}) \in \vec{i}(X^{t}_{i})$ ($i \neq j$). Non-local mutual information must be due to interactions because the initial states are independent (all pairs of units have zero mutual information). Finally we define the `integration' of information as the difference $I(X^{t}_{i} : X^{0}) - \sum_{j=1}^{N} I(X^{t}_{i} : X^{0}_{j})$. Information integration is not itself a member of $\vec{i}(X^{t}_{i})$ but it is fully induced by $\vec{i}(X^{t}_{i})$ since each of its terms is in $\vec{i}(X^{t}_{i})$. Therefore we will treat integration features as separate features in our results analysis but we do not add them to $\vec{i}(X^{t}_{i})$. 

\subsection*{Predicting the class of dynamical behavior using information processing features.}

\paragraph{Behavioral class of a rule.} Wolfram observed empirically that each rule $r \in 0,...,255$ tends to evolve from a random initial state to one of only four different classes of dynamical behavior~\cite{wolfram_new_2002}. These \emph{de facto} established behavioral classes are: 

\begin{enumerate}
\item Homogeneous (all cells end up in the same state);
\item Periodic (a small cycle of repeating patterns);
\item Chaotic (pseudo-random patterns); and
\item Complex (locally stable behavior and long-range interactions among patterns). 
\end{enumerate}

These classes are conventionally numbered 1 through 4 respectively. We obtained the class number for all 256 rules from Wolfram Alpha~\cite{wolframalpha} and denote it $C_r \in 1,...,4$. 

\paragraph{Predictive power of the information processing features.} At time step $t$ we numerically compute the sequence of information features $$\vec{I}^{t}_{r} \equiv \left(\vec{i}(X^{t=0}_{i}), \ldots, \vec{i}(X^{t}_{i})\right)$$ for rule $r \in \{0, 1, \ldots, 255\}$ to characterize its information processing up to time $t$. Then we formalize the prediction problem by the conditional probabilities $\Pr\left(C_{r} \mid \vec{I}^{t}_{r}\right)$, treating $r$ as an unobservable, uniformly random stochastic variable. That is, given only the sequence of information features of a rule, what is the probability that the ECA will eventually exhibit behavior of class $r$? We can interpret this problem as a communication channel $\vec{I}^{t}_{r} \rightarrow C_{r}$ and quantify the \textit{predictive power} of $\vec{I}^{t}_{r}$ using the mutual information $I(\vec{I}^{t}_{r} : C_{r})$. The predictive power is thus zero in case the information features do not reduce the uncertainty about $C_r$, whereas it achieves its maximum value $H(C_{r})$ in case a sequence of information features $\vec{I}^{t}_{r}$ always uniquely identifies the behavioral class $C_{r}$. We will normalize the predictive power as $I(\vec{I}^{t}_{r} : C_{r}) / H(C_r)$.

Note that a normalized predictive power of, say, $0.75$ does not mean that $75\%$ of the rules can be correctly classified. Our definition yields merely a relative measure where $0$ means zero predictive power, $1$ means perfect prediction, and intermediate values are ordered such that a higher value implies that a more accurate classification algorithm could in principle be constructed. The benefit of our definition based on mutual information is that it does not depend on a specific classifier algorithm, i.e., it is model-free. Indeed, the use of mutual information as a predictor of classification accuracy has become the \textit{de facto} standard in machine learning applications~\cite{battiti1994using,chow2005estimating}.


\paragraph{Selecting the principal features.} Some information features are more predictive than others for determining the behavioral class of a rule. Therefore we perform a feature selection process at each time $t$ to find these 'principal features' as follows. First we extend the set of information features $\vec{I}^{t}_{r}$ by the following set of `information integration' (also called `whole-minus-sum' (WMS)~\cite{griffith_synergy_entropy_2015}) features:


\begin{equation}
\vec{S}^{t}_{r} \equiv \{ I \left( X^{t'}_{i} : s \right) - \sum_{s_{i} \in s} I \left( X^{t'}_{i} : s_{i} \right)  : s \in 2^{X^{t=0}}, t' \in 0,...,t \}
\label{eq:wms}
\end{equation}

Their concatenation makes the extended feature set:

\begin{equation}
\vec{F}^{t}_{r} \equiv \left(\vec{I}^{t}_{r}, S^{t}_{r}\right) .
\end{equation}


The extended feature set $\vec{F}$ has no additional predictive power compared to $\vec{I}$ so for the prediction task $\vec{F}$ and $\vec{I}$ are equivalent. Namely, the integration features $\vec{S}^{t}_{r}$ are completely redundant given $\vec{I}^{t}_{r}$ since each of its terms is a member of $\vec{I}^{t}_{r}$. The reason for adding them separately is that they have a clear meaning as `information integration'~\cite{williams_nonnegative_2010,lizier_towards_2013,beer2015information,schneidman_synergy_2003,quax_synergy} and we are interested to see whether this phenomenon plays a significant role in generating dynamical behaviors. To illustrate this meaning, consider the first time step of rule $105$. At $t=1$ each cell computes its next state deterministically from its three predecessor states, transferring information from these states and indeed $I(X^{1}_{i} : X^{0}_{i-1}, X^{0}_{i}, X^{0}_{i+1})=1$. However, if we try to determine from which predecessor cells this information originated we find $\forall j: I(X^{0}_{i} : X^{0}_{j})=0$, i.e., none of the individual predecessor cells correlates with the cell's new state. This apparent paradox is resolved if we look more closely at the state transition table for this rule. Namely, the new state stores whether the sum of the predecessor states is even (0) or uneven (1), not whether one particular predecessor state is 0 or 1. This is a so-called `synergistic', higher-order relation which cannot be reduced to a sum of individual relations. Therefore we can say that the new unit state `integrates' three bits of information into one new bit of information. Other rules may partially integrate information or not at all. For our case of statistical independence among the input variables, the WMS functions precisely quantify this notion of information integration; in the general case it remains an open question how to quantify it.

We define the first principal feature $f^{t}_{1}$ as maximizing its individual predictive power, quantified by a mutual information term as explained before, as

\begin{equation}
f^{t}_{1} \equiv \argmax_{f_r \in \vec{F}^{t}_{r}} I\left(f_r : C_{r}\right).
\end{equation}

Here, $r$ is treated as a uniformly random stochastic variable with $\Pr(r)=1/256$ which in turn makes $f_r$ and $C_r$ stochastic variables. In words, $f^{t}_{1}$ is the single most predictive information feature about the behavioral class that will eventually be generated. More generally, the principal set of $n$ features is identified in similar spirit, namely

\begin{equation}
f^{t}_{n} \equiv \argmax_{\vec{f}_r \subseteq \left(\vec{F}^{t}_{r}\right)^n} I\left(\vec{f}_r : C_{r}\right).
\end{equation}




\paragraph*{Informational non-locality.} $C_{r}$ is a function of the long-term (large $t$) behavior of the system, after the transient phase $X^0 \mapsto X^1 \mapsto \ldots X^T$ from the i.i.d. initial state, so we can write loosely $C_r = f\left(X^T, X^{T+1}, \ldots, X^{T+M}\right)$. This means that the total predictive power $I(\vec{I}^{t}_{r} : C_{r})$ increases with time $t$ (as $t$ approaches $T$) but must converge at some $t=t_c$ to a maximum value, which is at most $H(C_{r})$ in case $C_r$ can be perfectly predicted from the information features. We refer to $t_c$ as the `informational non-locality'. This number reflects how many steps of information processing are needed to optimally predict which class of behavior will be generated, starting from a random initial configuration. It is a function of both the set of dynamical systems under investigation (in our case the 256 ECA rules) as well as the criterion for distinguishing dynamical behaviors (in our case the long-term behavior classification by Wolfram). 

To illustrate the concept in the general case, 
$t_c=1$ would imply that all interaction network topologies in the set of dynamical systems lead to the same behavioral class. Namely, in the first time step each unit operates on i.i.d. input states regardless of how their interactions are placed. Therefore $t_c=1$ implies that the placement of interactions is completely irrelevant. In contrast, $t_c=2$ would imply that the local network topology plays a role, since different interaction networks generate different correlated states at $t=1$ on which the unit process at $t=2$. Namely, in the second time step each unit operates on correlated states at distance 1 in the interaction network. In this case, changing a non-local network characteristic (such as the betweenness-centrality \cite{newman_structure_2003}) would not change the behavioral class that is generated, as long as the local network topological features (such as the transitivity coefficient \cite{newman_structure_2003}) are left invariant. The higher $t_c$, the larger the scale of the network features that play a role in generating the behavioral class. 

For the case of the 256 ECA the interaction network changes only locally, namely, implicitly by the choice of rule number: some rules ignore their left and/or right neighbor and thus effectively reduce or drop local pairwise (directed) interactions. In addition, the type of interactions can range from strictly one-to-one (e.g., copying the left neighbor's state) to strictly high-order `hyperedges'~\cite{newman_structure_2003} (e.g., rule 105). However, globally the interactions in the 256 ECA invariably form a homogeneous linear network, i.e., no global network properties are ever changed while keeping the local network properties equal. This means that the network structure can be learned already by any cell by looking at its neighbors and how its new state is generated from the neighbor states, as this will tell which interactions are dropped and which interactions are one-to-one or higher-order. It is possible that the interaction structure changes in the presence of correlations among the neighbors, in which case one or a few extra time steps will be needed before the eventual network is learned. Therefore we intuitively expect the $t_c$ for ECA to be rather small, on the order of 2 or 3, since after this many time steps the information feature values will (implicitly) capture all the network's variability. This hypothesis will be tested in the Results (Section~\ref{sec:results}).

\paragraph*{Information-based classification of rules.} The fact that Wolfram's classification relies on the behavior exhibited by a particular initial configuration makes the complexity class of an automaton dependent on the initial condition. Moreover, there is no universal agreement regarding how ``complexity'' should be defined and various alternatives to Wolfram's classification have been proposed, although Wolfram's remains by far the most popular. Our hypothesis is that the complexity of a system has very much to do with the way it processes information. Therefore we attempt to classifying ECA rules using only their informational features.

We use a classification algorithm which takes as input the 256 vectors of information features and computes the Euclidean distance between these vectors. The two vectors nearest to each other are clustered together. Then the remaining nearest elements or clusters are clustered together. The distance between two clusters is defined as the distance between the most distant elements in each cluster. The result is a hierarchy of clusters with different distances which we visualize as a dendrogram.

\subsection*{Computing information processing features in foreign exchange time series}

In the previous section we define information processing features for the simplest (one-dimensional) model of discrete dynamical systems. In the second part of this paper we aim at investigating if information features can distinguish ``critical'' regimes in the real complex dynamical system of the foreign exchange market. Most importantly, we are interested in the behavior of the information features before, at, and after the start of the 2008 financial crisis, which is commonly taken to coincide with the bankruptcy of Lehman Brothers on September 15, 2008. We consider two types of time series datasets in which the dynamical variables can be interpreted to form a one-dimensional system in order to stay as close as possible to the ECA modeling approach.

The information features can then be computed as discussed above, except that each mutual information term is now estimated directly from the data. This estimation is performed within a sliding window of length $w$ up to time point $T$ which enables to see how the information measures evolve over time $T$. For instance, the memory of variable $X$ at time $T$ will be measured as $I\left(X^t : X^{t+1}\right)$ where the joint probability distribution $\Pr(X^t, X^{t+1})$ is estimated using only the data points $X^{T-w}, \ldots, X^T$. Details regarding the estimation procedure are given in the following paragraph.

\paragraph*{Estimating information processing features from the data}

The mutual information between two financial variables (time series) at time $t$ is estimated using the $k$-nearest-neighbor algorithm using the typical setting $k=3$~\cite{kraskov_estimating_2004}. This estimation is calculated using a sliding window of size $w$ leading up to and including time point $t$, after first detrending each time series using a Gaussian smoothing kernel of width $\sigma$. Both parameters $w$ and $\sigma$ are evaluated for robustness. $\sigma$ is selected to be about half the value for which pairs of time series become co-integrated at the 0.05 significance level. 

For calculating the integration measure we apply a correction which makes it distributed around zero. The reason is that the WMS measure of Eq.~\ref{eq:wms} assumes independence among the stochastic state variables $s_i$, which for the real data is taken to be the previous day's data instead of the initial state. When this assumption is violated it can become strongly negative and, more importantly, co-integrated with the memory and transfer features whose sum will then dominate the integration feature. We remedy this by rescaling the sum of the memory and transfer features which are subtracted in Eq.~\ref{eq:wms} to equal the average value of the total information (positive term in Eq.~\ref{eq:wms}). This rejects the co-integration null-hypothesis between total information and the subtracted term at the 0.05 significance level ($p \approx 0.002$). This results in the integration feature being distributed around zero and being independent of the sum of the other two features so that it may functionally be used as part of the feature space, however the value itself should not be trusted as quantifying precisely the notion `integration' or synergy.

A similar procedure to remove the co-integration between memory and transfer is not possible because it would amount to having to assign the common dynamics to one of the features and making the other features only residuals. This means that this arbitrary choice would have a determining and potentially misleading impact on the distribution of the time points in the feature space.


\paragraph*{Description of the foreign exchange data}



The first data we consider are time series of five foreign exchange (FX) daily closing rates (EUR/USD, USD/JPY, JPY/GBP, GBP/CHF, CHF/EUR) for the period 1999-01-01 through 2017-04-21.\footnote{\url{http://www.global-view.com/forex-trading-tools/forex-history/}} Each currency pair has a causal dependence on its direct neighbors in the order listed because they share a common currency. For instance, if the EUR/USD rate changes then USD/JPY will quickly adjust accordingly (among others) because the rate imbalance can be structurally exploited for profit. In turn, among others through the rate JPY/EUR (not observed in this dataset) the rate EUR/USD will also be adjusted due to profit-making imbalances, eventually leading to all neighboring rates returning to a balanced situation.

\paragraph*{Description of the interest-rate swap data}

The second data are interest-rate swap (IRS) daily rates for the EUR and USD market \cite{quax2013information}. The data spans over twelve years: the EUR data from 1998-01-12 to 2011-08-12 and the USD data from 1999-29-04 to 2011-06-06. The datasets consist of 14 and 15 times to maturity (durations), respectively, ranging from 1 year to 30 years. Rates for nearby maturities have a dependency because the higher maturity can be constructed by the lower maturity plus a delayed (`forward') short-term swap. This basic mechanism between maturities leads to generally monotonically upward `swap curves'.

\section*{Results}
\label{sec:results}

\subsection*{Predicting the Wolfram class of ECA rules using information processing features}

\begin{figure}[t]
\centering
\includegraphics[width=0.85\textwidth]{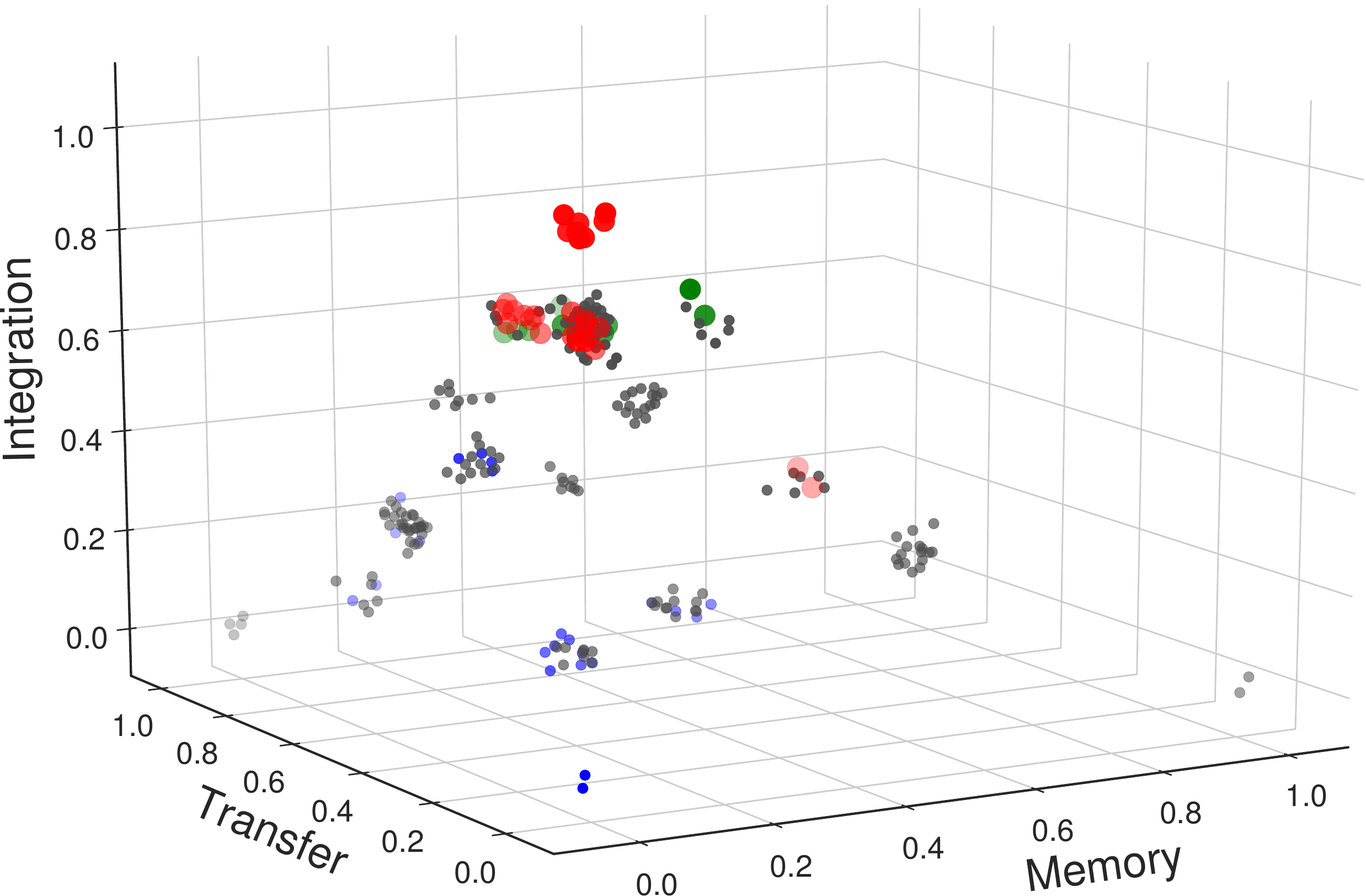}
\vspace{0.5cm}
\caption{Three selected features from $F^{t=1}_r$ for each ECA rule $r \in 0..255$, namely $I(X^{t=1}_{i} : X^{t=0}_{i})$ ('memory'), $I(X^{t=1}_{i} : X^{t=0}_{i-1}) + I(X^{t=1}_{i} : X^{t=0}_{i+1})$ ('transfer'), and $I(X^{t=1}_{i} : X^{t=0}_{i-1}, X^{t=0}_{i}, X^{t=0}_{i+1}) - \sum_{\delta \in \{-1,0,+1\}} I(X^{t=1}_{i} : X^{t=0}_{i+\delta})$ ('integration'). Each dot corresponds to a rule $r$ and is color-coded by their Wolfram class $C_r$, namely black for the simple homogeneous (24) and periodic behaviors (196), green for complex behavior (10), and red for chaotic behavior (26). The transparency of a point indicates its distance away from the viewer, with more transparent points being farther away. A small random vector with average norm $0.02$ is added to each point in order to make rules with equal information features still visible.}
\label{fig:cube}
\end{figure}

\paragraph{Information processing in the first time step.} The information processing occurring in the first time step of each ECA rule $r \in 0..255$ is characterized by the corresponding feature set $\vec{F}^{t=1}_r$, consisting of 7 time-delayed mutual information quantities ($\vec{I}^{t=1}_{r}$) and 4 information integration quantities ($\vec{S}^{t=1}_{r}$). We show three selected features for all 256 rules as points in a vector-space in Fig.~\ref{fig:cube} along with each rule's Wolfram class as a color code. It is apparent that the three features already partially separate the behavioral classes. Namely, it turns out that chaotic and complex rules tend to have high information integration, low information memory, and low information transfer. Fig.~\ref{fig:cube} also relates intuitively to the classic categorization problem in machine learning, namely, perfect prediction would be equivalent to the existence of hyperplanes that separate all four behavior classes.


\begin{figure}[h!]
\centering

\includegraphics{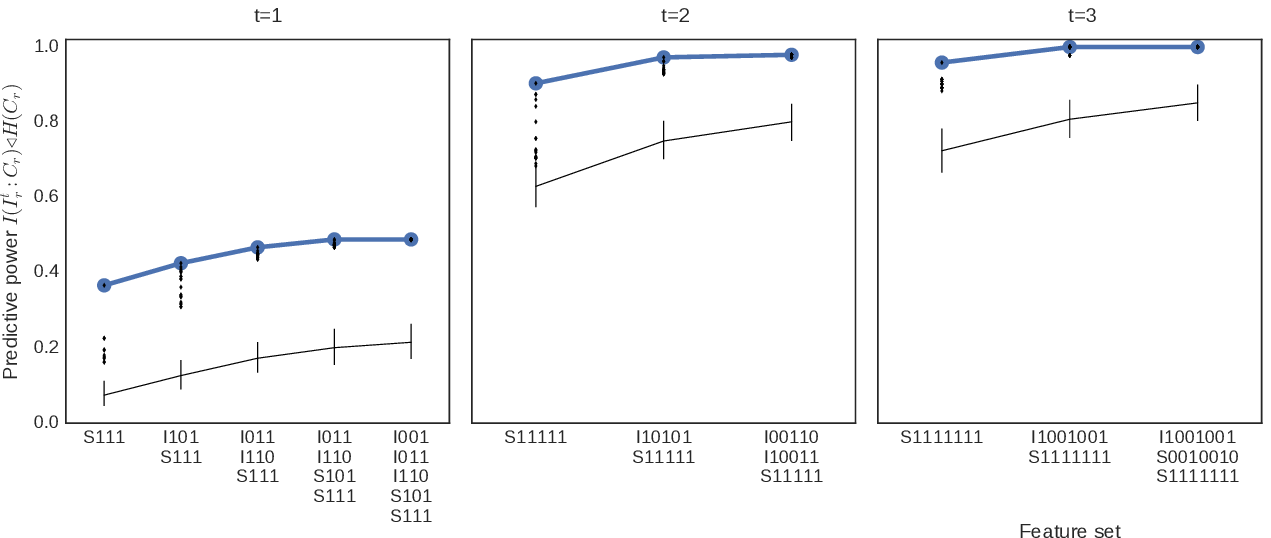}
\caption{Predictive power of the optimal set of $n$ information features as function of $n$ (thick blue line with round markers). For each $n$ the best set of information features is listed: \texttt{S} means information integration ('synergy') and \texttt{I} means mutual information, followed by a bitmask of which neighbor cell's initial state is included in the feature (middle bit is the cell itself). For example, \texttt{I010} indicates the `memory' feature $I(X^1_i : X^0_i)$, and \texttt{S111} denotes $I(X^{t=1}_{i} : X^{t=0}) - \sum_{j} I(X^{t=1}_{i} : X^{t=0}_{j})$. The thin black line with error bars is the `base line' predictability for $n$ features, obtained by randomizing the pairing of 256 information features with their class identifier. The error bar indicates the 95\% confidence interval of the distribution of predictive power under the null-hypothesis of zero correlation between information features and class identifier. Finally, the small black markers indicate the predictive powers of all other information feature sets.}

\label{fig:predictive_power}
\end{figure}

\paragraph{Predictive power of information processing features over time.} The single most predictive information feature is \emph{information integration}, as shown in Fig.~\ref{fig:predictive_power}. Its predictive power is about $0.37$ where $1.0$ would mean perfect prediction. The most predictive \emph{pair} of features is formed by adding information \emph{transfer} at $0.43$, so adding the information transfer feature increases the predictive power by $0.06$. The information transfer feature by itself has actually over three times this predictive power at $0.19$, showing that two or more features can significantly overlap in how they characterize the behavioral class. The total predictive power of all information processing features at $t=1$ is $0.49$, formed by 4 of the 11 possible information features.

For the second time step (Fig.~\ref{fig:predictive_power}) we again find that the most predictive information feature is information integration. An intriguing difference however is that it is now significantly more predictive at $0.90$. This means that already at $t=2$ there is a single information characteristic of dynamical behavior (i.e., information integration) which explains the vast majority of the entropy $H\left(C_r\right)$ of the behavioral class that will eventually be exhibited. A second intriguing difference is that the maximum predictive power of $0.98$ is now achieved using only 3 out of 57 possible information features, where 4 features were needed at $t=1$.

Finally, for $t=3$ we find that only 2 information features achieve the same possible maximum predictive power of $1.0$, i.e., the value for these two features unique identify the behavior class. Firstly this confirms the apparent trend that fewer information features capture more of the relevant dynamical behavior as time $t$ increases. Secondly we find again that information integration is the single most predictive feature. In addition, we find again that the best secondary feature is a particular combination of memory and longest-range transfers, as in $t=2$. Including the intermediate transfers actually only slightly convolutes the prediction: adding them in $t=2$ reduces predictive power by $0.028$, whereas in $t=3$ only by $4\cdot 10^{-15}$. In $t=1$ there are no intermediate transfers possible since there are only three predecessors of a cell's state, and apparently then it pays off to leave out memory (which would reduce power by $0.025$ if added).

To validate that the predictive power values of the information features are indeed meaningful we also plot the expected `base line' prediction power in each panel in Fig.~\ref{fig:predictive_power} along with the 95\% confidence interval. The base line is formed by randomizing the pairing of information features with class identifiers, so it shows the expected predictive power of having the same number and frequencies of feature values but sampled with zero correlation with the classification. In effect this results in a statistical test with a null-hypothesis of the information features and Wolfram classification being uncorrelated. Since the predictive power of the information features are always above the base line we consider the meaningfulness of the information features validated, i.e., we reject the null-hypothesis at the 95\% confidence level.

\paragraph{Informational non-locality.} In the first time step each unit operates on random and independent states by our restriction of the initial conditions, so the system state $X^{t=1}$ is invariant under the network topology of the interaction. That is, regardless of whether the interaction network is a line graph or binary tree graph, the stochastic variable $X^{t=1}$ has the same distribution and is thus insensitive to the interaction topology. Only the number of interactions (degree) per cell may have an effect at this stage, but for ECA it is invariably 2. We can thus say that the information features $\vec{I}^{t=1}_{r}$ only characterize the local unit dynamics (i.e., their rule table), but does not yet take into account which unit interacts with which other units. The total predictability at $t=1$ of $0.49$ thus suggests that the local unit dynamics alone play already a significant role in determining the behavioral class that will be generated, almost on equal footing with the network topology of the interactions among the units, which should then account for the remaining $0.56$. 

In the second time step each unit operates on possibly correlated states, induced by the interaction topology, where two correlated states are at most distance 2 away from each other. The system state $X^{t}$ thus depends on highly local network characteristics, such as whether the two neighbors of a unit are also connected with each other or not. That is, a line graph and a sequence of triangles will still yield the same $X^{t}$, but a binary tree graph would yield a different $X^{t}$. We find that the total predictability at $t=2$ equals $0.98$, so adding this dependence on highly local network characteristics increases the predictability by $0.49$.

In the third time step and beyond we find that the total predictive power reaches $1.0$, so we find that the `informational non-locality' equals $t_c=2$. This finite value means that system-size network characteristics, such as average betweenness centrality \cite{newman_structure_2003}, play no role in determining the behavioral class in the set of 256 ECA rules. Indeed this is to be expected since large-scale network characteristics do not change across the 256 ECA rules: globally they are all line graphs of infinite length. Only at the smallest scale the interaction networks vary effectively since some rules (partly) ignore their `left' neighbor's state, whereas others ignore their `right' neighbor's state, or both. This explains why only the network characteristics at the smallest scale play a role in determining the behavioral class of ECA rules, which is successfully recovered by our $t_c$ measure. We expect that for more heterogeneous classes of dynamical systems, such as random boolean networks, the informational non-locality is (much) higher, but we leave this as a future study.

\paragraph{Relation to Langton's parameter.} Langton's $\lambda$ parameter~\cite{langton1990computation} is the most well-known single feature of an ECA rule which partially predicts the Wolfram class. It is a single scalar computed for rule $r$ as $\lambda_r = \Pr(X^{1}_{i} = 0 \mid r)$. It is known that the $\lambda$ parameter is more effective for a larger state space and a larger number of interactions, however we briefly highlight it here because of its widespread familiarity and because the information processing measures can be written in terms of `generalized' $\lambda$ parameters (see Supporting Material). This means that $\lambda$'s relation with the Wolfram classification is captured within the information features, implying that the information are minimally as predictive as features based on $\lambda$ parameter(s). 

Indeed, the predictive power $I(\lambda : C_r)/H(C_r) \approx 0.175$, which is significantly lower than the information integration feature alone which achieves $0.361$ at $t=1$. Moreover, as indicated by the black dots in the left panel of Fig.~\ref{fig:predictive_power} the vast majority of information features have higher predictive power than $0.175$ (only three single features have slightly lower power).

\paragraph{Information processing-based clustering.} In the previous sections we showed how information features predict Wolfram's behavioral classification. In this section we investigate the hierarchical clustering of rules induced by the information features in their own right. One important reason for studying this is the fundamental problem of undecidability of CA classifications based on long-term behavior characteristics \cite{culik1988undecidability,sutner2012classify}, such as for the Wolfram classification. In the best case this makes the classification difficult to compute; in the worst case the classification is impossible to compute, leading to questions about its meaning. In contrast, if local information features correlate strongly with long-term emergent behaviors then a CA classification scheme based on information features is practically more feasible. In this section we visualize how the clustering overlaps with Wolfram's classification.

\begin{figure}[t]
\centering
\includegraphics[width=0.95\textwidth]{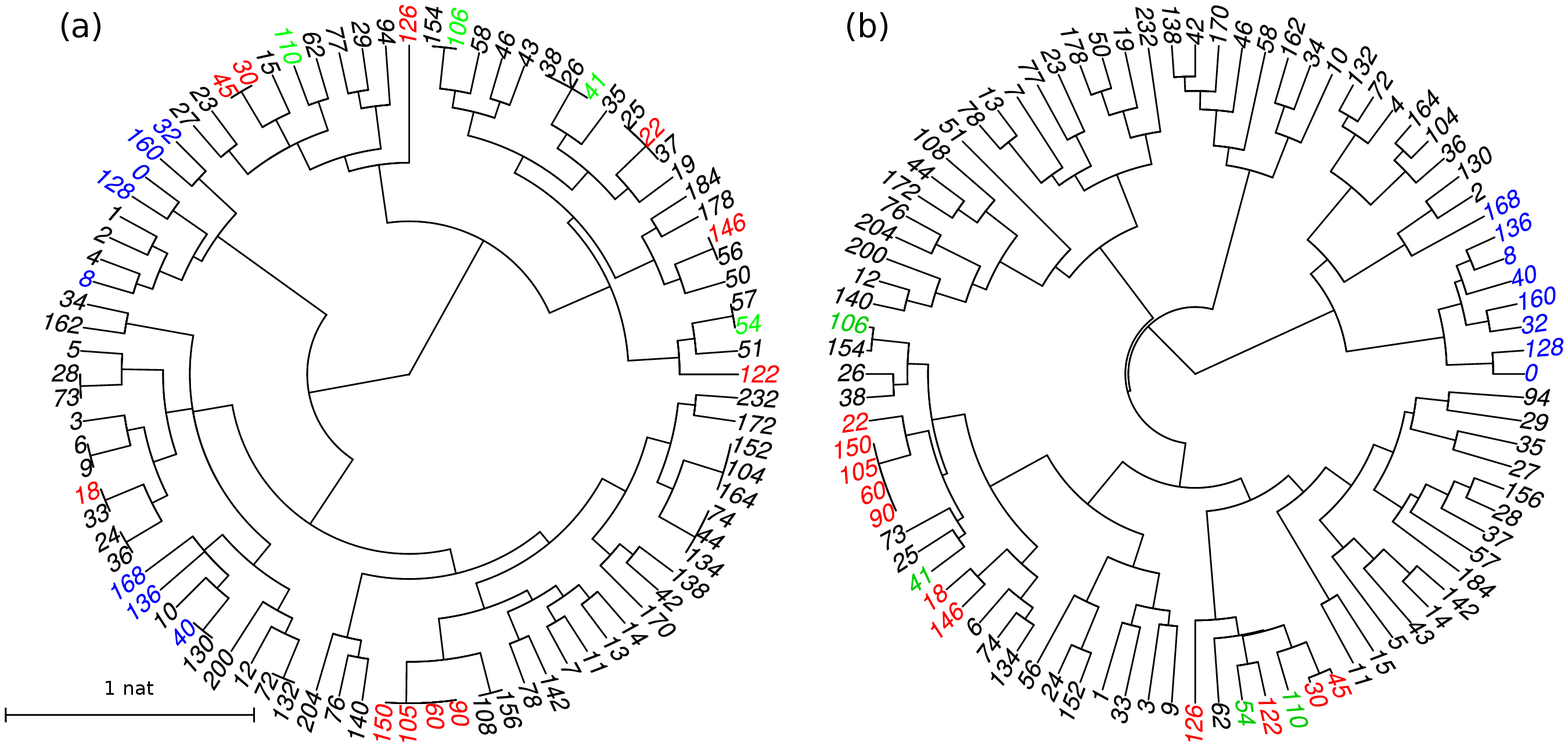}
\caption{\emph{Dendrograms displaying two clusterings of information-processing features: (a) when features are evaluated when units are initially uncorrelated, and (b) when features are evaluated numerically when cells are in their equilibrium state. The colour code is the same as in Fig.~\ref{fig:cube} except that rules in class I are displayed in blue to distinguish them from class II.}}
\label{fig:4tree}
\label{fig:8tree}
\end{figure}

Fig. \ref{fig:4tree} (a) shows a clustering made using information features evaluated where $t=0$ is the randomized state. Interestingly, while the features have a low predictability on the Wolfram class, the resulting clustering also does not overlap at all with Wolfram classification. We have to make an exception for rules 60, 90, 105 and 150 which are all chaotic rules and share the same information processing features.

Fig. \ref{fig:8tree} (b), on the other hand, displays the clustering for the case where features are not evaluated with respect to the randomized state but to the stationary distribution (i.e., $X^{t=0}$ is a randomly selected system state from the cyclic attractor). One reason for this is to ignore the initial transient phase; another reason is to make a step towards the financial application in the next subsection, which obviously does not start from random initial conditions. By `stationary' we mean that we simulate the CA dynamics long enough until the time-shifted information features no longer have a changing trend. Feature values can no longer be calculated exactly and thus are estimated numerically by sampling initial conditions and for size $N=15$. In that case we find that the clustering has increased overlap with Wolfram's classification. In particular, we can note that uniform rules cluster together and that chaotic and complex rules are all on the same large spray. However the agreement is far from perfect. For instance, the spray bearing chaotic and complex rules also bears periodic rules. Note also that rules 60, 90, 105 and 150  are indistinguishable when considered from the information processing viewpoint, even though they exhibit chaotic patterns that can be visually distinguished from each other. On the contrary, rules 106 and 154 are very close to each other and the pattern they exhibit indeed shows some similarities, but the former is complex while the latter is periodic.

Note that using this clustering scheme all rules converging to a uniform pattern, but one, are close to each other in the information-features space. The remaining one, rule 168, has a transient regime which is essentially dominated by a translation of the initial state. This translational regime can be found as well in rules 2 and 130, which are classified in the same sub-spray as rule 168. The similarity of any information feature (information transfer in this case) can thus lead to rules whose behavior differs in other respects to get classified similarly.

\subsection*{Detecting a regime shift in foreign exchange time series}

Inspired by the results for the ECA models we study here whether a small set of information features is also capable of identifying different `regimes' of dynamical behavior of real systems using real data. We focus on financial data because it is of high quality and available in large quantities compared to for instance biological datasets. Also at least one large regime shift is known: the 2008 financial crisis, separating the pre-crisis and post-crisis regimes. We focus on two time series datasets: daily IRS rates of 14 and 15 maturities in the USD and EUR markets, and five consecutive daily FX closing exchange rates. We selected these datasets because the variables in each dataset can be considered to form a line graph similar to ECA rules, staying as close as possible to the previous analysis.

\begin{figure}
\includegraphics[width=1.0\textwidth]{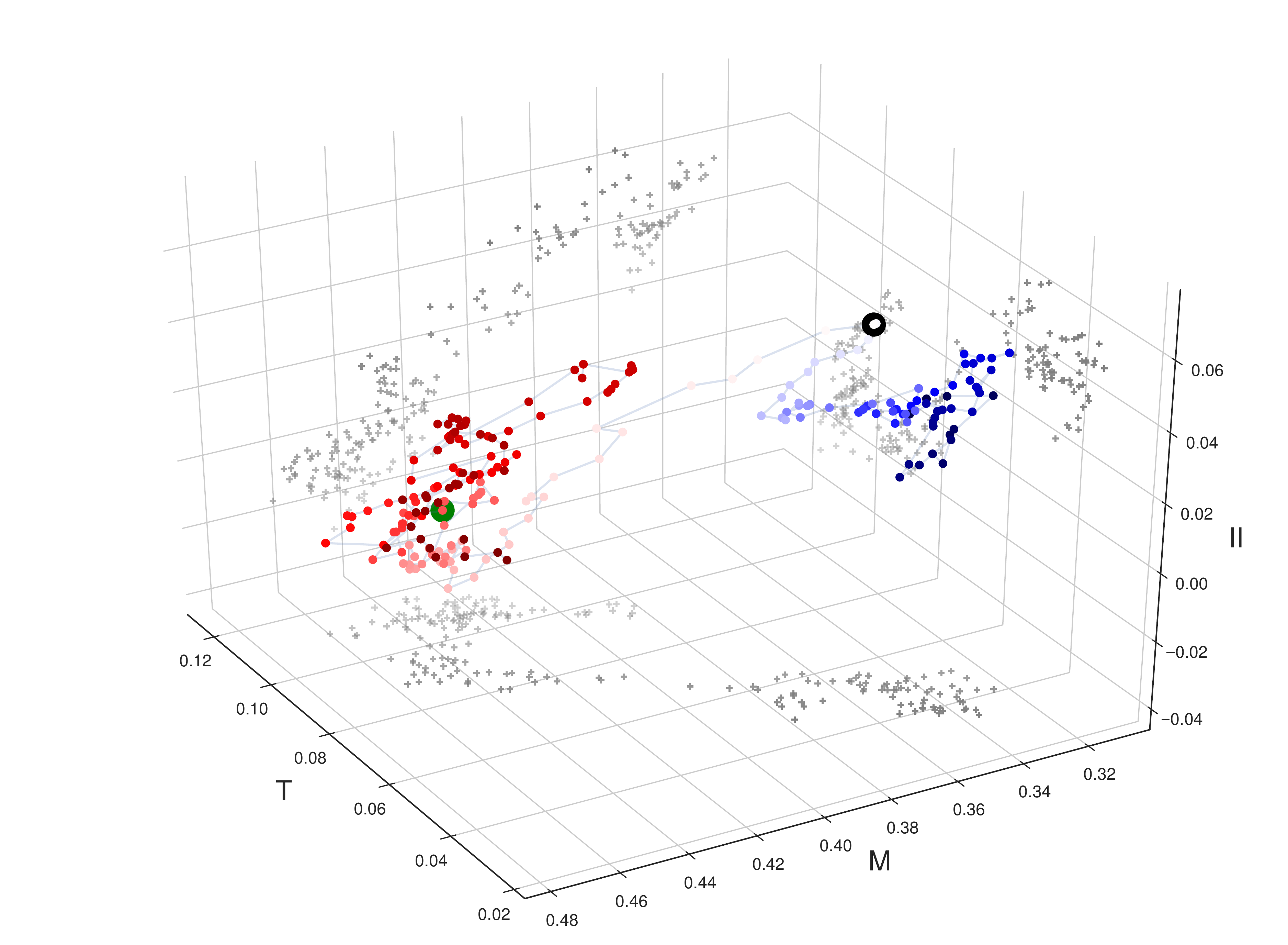}
\caption{200 time points showing the progression of the three information features memory (M), transfer (T), and integration (II) computed with a time delay of 1 day (similar to $t=1$ for ECA). The color indicates the time difference with September 15, 2008 (big black dot), which we consider the starting point of the 2008 crisis, from dark blue (long before) to dark red (long after) and white at the crisis date. The data spans from 1999-01-01 through 2017-04-21; the large green dot is the last time point also present in the IRS data in 2011. In this information space we clearly observe signs of a two attractor regimes separated by a sudden regime shift. The shift is preceded by a slow directed, non-stationary progression in the pre-crisis (blue) regime, resembling the dynamics of a tipping point~\cite{scheffer_early-warning_2009}. Mutual information is calculated using a sliding window of $w=1400$ days.}
\label{fig:3dscatter_fx}
\end{figure}

In Figure~\ref{fig:3dscatter_fx} we show the 3-dimensional `information feature space' with the same axes as Figure~\ref{fig:cube}. We observe a remarkable separation of the pre-crisis and post-crisis periods which are well separated by a fast transition trajectory. Before this transition we also observe in the pre-crisis regime that the information features are not stationary but slowly progress along a path in the general direction of the post-crisis regime (high memory, high transfer, low integration) before the fast transition. Interestingly, this closely resembles the dynamics observed for so-called tipping points~\cite{scheffer_early-warning_2009} where a system is slowly pushed `over the hill' after which it progresses quickly `downhill' to the next attractor state. This is relevant because slow progressions to tipping points offer a potential for developing early-warning signals~\cite{scheffer_early-warning_2009}, however we do not further explore this possibility here.

\subsection*{Detecting a regime shift in interest-rate swap time series}

\begin{figure}
\centering
\includegraphics[width=0.9\textwidth]{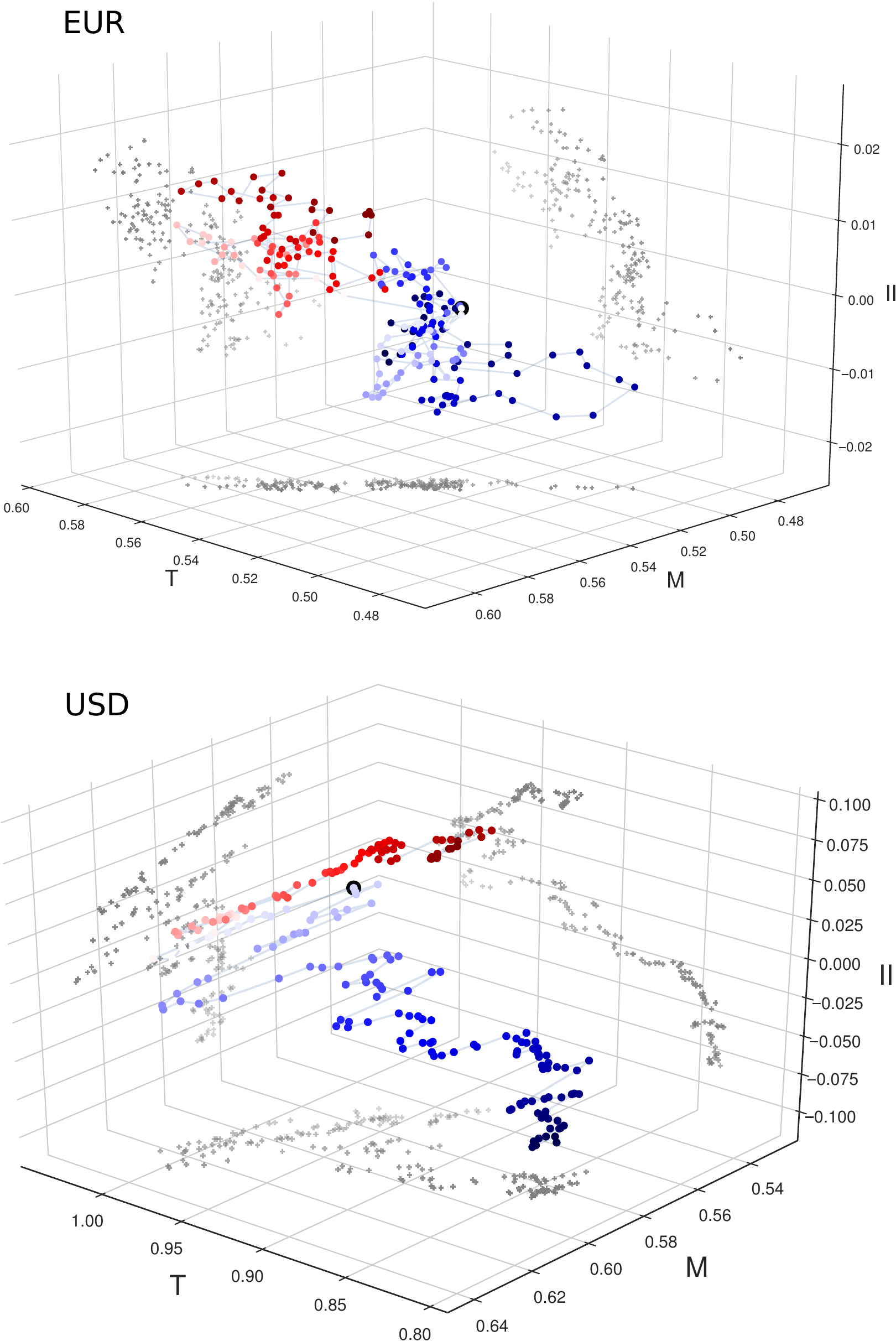}
\caption{200 time points showing the progression of the three information features memory (M), transfer (T), and integration (II) computed with a time delay of 1 day (similar to $t=1$ for ECA). The color indicates the time difference with September 15, 2008 (big black dot), which we consider the starting point of the 2008 crisis, from dark blue (long before) to dark red (long after) and white at the crisis date. The data spans more than twelve years: the EUR data from 1998-01-12 to 2011-08-12 and the USD data from 1999-29-04 to 2011-06-06. In the EUR market two regimes can be identified which are well-separated but close together; in the USD market we see instead a steady, prolonged progression in a similar direction. Mutual information is calculated using a sliding window of $w=1400$ days.}
\label{fig:3dscatter_irs}
\end{figure}

In Figure~\ref{fig:3dscatter_irs} we show the same feature space for the IRS markets in EUR and USD. In EUR we similarly see a good separation of pre-crisis and post-crisis into two separate regimes, although the two regimes are closer together compared to the FX case. A slow progression towards a tipping point is less clear in this case.

In contrast, in USD we observe the completely different scenario of a steady progression of the information features during most of the duration of the dataset. One possible but hypothetical explanation for this is that the IRS market in USD could have been (part of) a slow but steady driving factor in the global progression to the crisis, whereas the EUR IRS and the FX markets may have been more exogenously forced towards their regime shift. Indeed, the progression to the 2008 crisis is often explained by referring at least to substantial losses in fixed income and equity portfolios followed by the U.S. subprime home loan turmoil~\cite{Melvin20091317}, suggesting at least a central role for the trade of risks concerning interest rates in USD. The exact sequence of events leading to the 2008 crisis is still debated among financial experts and our numerical analyses may help to shed light on interpreting the relative roles of different markets.

Another intriguing observation is that financial markets appear to settle to a new steady state after the crisis as opposed to reverting back to their original dynamics. This would suggest that there is no `in and out of' the financial crisis; instead, the `crisis' appears to be actually a substantial and sustained regime shift. This is corroborated by a series of new, post-2008 market practices. A crucial development is that before 2008 counterparty credit risk was largely ignored whereas since 2008 it is explicitly priced in various forms~\cite{saunders2010credit}. For instance, it has become market standard to charge credit valuation adjustments (CVA) for unsecured over-the-counter derivative trades. CVA is the difference between the risk-free portfolio value and the true portfolio market value that takes into account the possibility of a counterparty’s default. Although research into how exactly to calculate CVA is still ongoing~\cite{graaf2014cva}, it is clear that the post-crisis systemic market dynamics have markedly and sustainably changed, especially regarding the detection and mitigation of various forms of counterparty and systemic risks.

Nevertheless at this stage it remains unclear why the information processing features move towards higher memory, higher transfer, and lower integration (FX) or higher integration (IRS). We note that the memory and transfer features can be reliably estimated from the data and that their relative change is consistent with a move from complex and chaotic ECA rules toward class 1 and 2 rules. We also note that memory and transfer are correlated in the real data, leading to the diagonal tendency of the time points in the feature spaces. The integration feature must be considered with more caution since the existence of (strong) correlations makes it problematic to exactly quantify it. Still it does become clear that this feature plays a significant role.

A possible explanation for the higher integration in the IRS markets may be that the rates are increasingly determined by counterparty and systemic risk indicators. After all these indicators are surely aggregate functions of IRS rates (among others) as opposed to direct transformations of an individual IRS rate, leading to reduced individual correlations and thus to higher integration. Risk indicators play a much less significant role in the dependencies among foreign exchange rates, offering a possible explanation for the observation that its integration at least does not increase. Although the absolute exchange rates may depend on external systemic risk indicators, leading to a larger exogenous factor in determining the rates, our measures would not detect this because mutual information is insensitive to monotonic transformations of the variables. The way in which different rates depend on each other remains largely the same: the FX market is a rather technical market with a large fraction of algorithmic traders which exploit (small) imbalances between different rates. Also, transactions are instantaneous as opposed to the long-term IRS contracts, meaning no counterpart dependencies and thus no risk factors incorporated in the way that different rates depend on each other.

\section*{Discussion}
Our working assumption is that dynamical systems inherently process information. Our leading hypothesis is that the way that information is locally processed determines the global emergent behavior. In this article we formalize the notion of `information processing' and then present two lines of supporting evidence: from a model perspective (ECA) and a data-analysis perspective (FX data). 

Our formalization builds upon Shannon's information theory, which means that we consider an ensemble of state trajectories rather than a single trajectory. That is, we do not quantify the information processing that occurs during a particular, single sequence of system states (attempts to this end are due to Lizier et al. \cite{lizier_local_2012}). Rather, we consider the ensemble of all possible state sequences along with their probabilities. One way to interpret this is that we quantify the `expected' information processing averaged over all trajectories \cite{lizier_local_2012}. Another way to interpret it is that we characterize a dynamical \emph{model} in its totality, rather than a particular symbolic sequence of states of a model. Our reasoning is that if (almost) every state trajectory of a model (such as a CA rule) leads to a particular emergent behavior (such as chaotic or complex patterning) then we would argue that the emergent behavior is a function of the ensemble dynamics of the model.

This seems at odds with computing information features from real time series, which are measurements of a single trajectory of a system. We resolve this issue by assuming `local stationarity'. This assumption is common in time series analysis and used (implicitly or explicitly) in most `sliding window' approaches and moving statistic estimations, among others. In other words, we assume that the rate of sampling data points is significantly faster than the rate at which the underlying statistical property changes, which in our case are the information features. The consequence is that a finite number of consecutive data points can be used to estimate the probability distribution of the system at the corresponding time, which in turn enables estimating mutual information quantities. 

Our first intriguing result from the ECA analysis is that fewer information features capture more of the relevant dynamical behavior, as time $t$ progresses away from a randomized system state. One potential explanation is the processing of correlated states, or equivalently, of overlapping information. Namely, at $t=1$ each cell operates exclusively on uncorrelated inputs, so the resulting state distribution is a direct product of the state transition table. Neighboring cell states at $t=1$ are now correlated due to overlapping input states, induced by the interaction topology. Consequently, at time $t=2$ and beyond, the inputs to each cell have become correlated in a manner dictated by the interaction topology. The way in which an ECA rule subsequently deals with these correlations is evidently an important characteristic. In other words, two ECA rules may have exactly the same information features for $t=1$ but different features for $t=2$ which must be due to a different way of handling (combining) correlated states.

This result leads us to propose the \emph{information non-locality} concept quantified by $t_c$. Namely, the information measures at $t=1$ of each cell do not yet characterize the interaction topology (by the correlations it induces) other than perhaps the degree distribution. This suggests that the `missing' predictive power at $t=1$ is a measure of the relevance of the (non-local) interaction topology. In the case of ECA this quantity is roughly half: $1-0.488=0.512$. In other words, if no correlations would ever be induced at $t=1$ and the CA still always operates on a randomized state at $t=2$ then the information measures at $t=2$ would yield zero additional predictive power about the eventual emergent behavior, which would lead to $t_c=2$. To the extent that correlations are induced, this added predictive power increases. Indeed we find $t_c=3$ for ECA. This concept extends to further time steps at which correlations over larger distances could be induced. The distance \em or time step $t_c$ \em at which the predictive power no longer increases is a measure of the `locality' of topological characteristics that are relevant for the emergent behavior.

To illustrate the potential importance of $t_c$, consider the set of all possible gene-gene regulatory interaction networks as the universe of dynamical systems, and consider the stability of the generated patterns as behavioral classes. $t_c$ then reflects the relative importance of the network topology versus the local dynamics. In turn, this would also inform the experimentation and modeling: if a large $t_c$ is found then any modeling attempt must take great care to measure and reproduce the large-scale network characteristics beyond only the degree distribution, and conversely, a low $t_c$ may imply for instance that only the degree distribution and the clustering coefficient need to be taken into account. 

Our second intriguing result is that the most predictive information feature is invariably \emph{information integration}. In each time step it accounts for the vast majority of the total predictive power ($75\%$, $92\%$, and $96\%$, respectively). This is the feature that we would consider to actually capture the `processing' or modification of information, rather than the memory and transfer features which capture the simple `copying' of information. Indeed, the cube of Fig.~\ref{fig:cube} suggests that the interesting behaviors (chaotic and complex) are associated with high integration, low memory, and low transfer. In this extreme we find rule 60 (the XOR rule) and similar rules which are all chaotic rules. For complex behavior non-zero but low memory and transfer appear to be still necessary ingredients.



The good separation of the dynamic behavioral classes in the ECA models using only a few information features ultimately leads to the question whether the same can be done for real systems based on real data. This is arguably a large step and certainly more rigorous research should be done using intermediate models of increasing complexity and for different classifications of dynamical behavior. On the other hand, if promising results could be obtained from real data using a small set of information features then this would add more urgency to such future research, even if not yet fully understanding the role of information processing features in systemic behavior. This is the purpose of our application to financial data. Financial data is of high quality and available in large quantities and at least one large regime shift is known, namely the 2008 financial crisis. We stay as close as possible to ECA by selecting two datasets in which the dynamical variables can be interpreted to form a line graph. Indeed, we consider our results in the financial application promising enough to warrant further study into information processing features in complex system models and other real datasets. Our results suggest tipping point behavior for the FX and EUR IRS markets and a possible driving role for the USD IRS market.

All in all we conclude that the presented information processing concept appears indeed to be a very promising framework to study how dynamical systems generate emergent behaviors. In this paper we present initial results which support this claim. Further research may identify concrete links between information features and various types of emergent behaviors, as well as the relative impact of the interaction topology. Our lack of understanding of emergent behaviors is exhibited by the ECA model: it is arguably the simplest dynamical model possible, and the choice of local dynamics (rule) and initial conditions fully determine the emergent behavior that is eventually generated. Nevertheless even in this case no theory exists that predicts the latter from the former. The information processing concept may provide a new perspective on dynamical systems with potentially pervasive impact across the sciences.


\newpage
\section*{Supporting Information}

\subsection*{Lambda parameter and information features}

\paragraph{Relation to Lambda parameter.}

The Lambda parameter is a well-known `local' characterization of a CA rule which correlates with Wolfram's complexity classes. It is local in the sense that it is computed from the state transition table, so effectively it can be seen as a prediction of the Wolfram class at $t=0$. Here we demonstrate that our information features includes part of the predictive power of the Lambda parameter by relating the two.

Let us first consider a dynamical system with $N$ variables whose state $a_i(t)$ evolves in time according to a given rule, depending on the state of the neighborhood of agent $i$, denoted $n_i$ (we shall always assume the neighborhood of a node includes the node itself). We assume a probabilistic description so as to write this rule as a transition matrix

\begin{equation}
W_{\{a_j\} \to a_i} = P(a_i, t \vert \{a_j\}, t-1)
\end{equation}
$\forall j \in n_i$. This gives the probability that agent $i$ takes value $a_i$ at time $t$ knowing that the neighbors of $i$ take value $a_{j_k}$ for $k \in \{1, ..., \vert n_i \vert \}$. With the $W$ matrix we can express two-steps joint probabilities as

\begin{equation}
P(a_i, t;\{a_j\}, t-1) = P(\{a_j\}, t-1) W_{\{a_j\} \to a_i} .
\end{equation}

We can now define and compute the \emph{total information} $I_{tot}$ as

\begin{align}
I_{tot} & := I(a_i, t; \{a_j\}, t-1) \notag \\
& = \sum_{\{a_j\}, a_i} P(a_i, t;\{a_j\}, t-1) \ln \frac{P(a_i, t;\{a_j\}, t-1)}{P(a_i, t)P(\{a_j\}, t-1)} \notag \\
& = \sum_{\{a_j\}, a_i} P(\{a_j\}, t-1) W_{\{a_j\} \to a_i} \ln \frac{W_{\{a_j\} \to a_i}}{P(a_i, t)}
\end{align}
Now comes a crucial assumption: we shall assume that the agents are randomly initialized, namely that $P(\{a_k\}, t-1) = 2^{-\vert n_k \vert}$ for any configuration of the neighborhood. In the case of cellular one is also allowed to write $n_i = n$ $\forall i$, so that

\begin{align}
I_{tot} & = 2^{-n} \sum_{\{a_j\}, a_i} W_{\{a_j\} \to a_i} \ln \frac{W_{\{a_j\} \to a_i}}{2^{-n} \sum_{\{a_j\}} W_{\{a_j\} \to a_i}} \notag \\
& = 2^{-n} \sum_{\{a_j\}, a_i} W_{\{a_j\} \to a_i} \ln W_{\{a_j\} \to a_i} - 2^{-n} \sum_{\{a_j\}, a_i} W_{\{a_j\} \to a_i} \ln \left( 2^{-n} \sum_{\{a_j\}} W_{\{a_j\} \to a_i} \right) \notag \\
& = - 2^{-n} \sum_{\{a_j\}, a_i} W_{\{a_j\} \to a_i} \ln \left( 2^{-n} \sum_{\{a_j\}} W_{\{a_j\} \to a_i} \right) \notag \\
& = - \lambda \ln \lambda - (1-\lambda) \ln (1-\lambda)
\end{align}
where $\lambda$ is the Langton parameter, namely the fraction of ones in the lookup table characterizing the dynamics. The last equality arises from the fact that

\begin{equation}
\sum_{\{a_j\}} W_{\{a_j\}\to a_i}=\left\{
 \begin{array}{cc}
2^n\lambda & \mbox{if $a_i=1$}\\
2^n(1-\lambda) & \mbox{if $a_i=0$}\\
\end{array}
\right.
\end{equation}
while the penultimate is because for a deterministic dynamics transition elements take either value 0 or 1.

The next piece of information we consider is \emph{memory}, which we define as

\begin{equation}
I_{mem} := I(a_i, t; a_i, t-1) .
\end{equation}
It can be computed as

\begin{align}
I_{mem} & = \sum_{a_i, a'_i} P(a_i, t;a'_i, t-1) \ln \frac{P(a_i, t;a'_i, t-1)}{P(a_i, t)P(a'_i, t-1)} \notag \\
& = \sum_{a_i, \{a'_j\}} P(a_i, t;\{a'_j\}, t-1) \ln \frac{\sum_{\{a'_{j \neq i}\}} P(a_i, t;\{a'_j\}, t-1)}{P(a_i, t)P(a'_i, t-1)} \notag \\
& = 2^{-n} \sum_{a_i, \{a'_j\}} W_{\{a'_j\}\to a_i} \ln \frac{\sum_{\{a'_{j \neq i}\}} W_{\{a'_j\}\to a_i}}{2^{-1} \sum_{\{a'_j\}} W_{\{a'_j\}\to a_i}}
\end{align}
We now refine the previous analysis by defining two quantities $\lambda_0$ and $\lambda_1$ as Langton parameters restricted to the case when the central site is $0$ or $1$ respectively. In other words we have

\begin{equation}
\sum_{a'_m, m\ne i} W_{\{a'_m,a'_i\}\to a_i}=\left\{\begin{array}{cc}
   2^n (1/2-\lambda_0)& \mbox{if $a_i=0$ and $a'_i=0$}\\
   2^n \lambda_0    & \mbox{if $a_i=1$ and $a'_i=0$}\\
   2^n (1/2-\lambda_1)& \mbox{if $a_i=0$ and $a'_i=1$}\\
   2^n \lambda_1    & \mbox{if $a_i=1$ and $a'_i=1$}\\
  \end{array}\right.
\end{equation}
so that $\lambda_0 + \lambda_1 = \lambda$. This allows to rewrite $I_{mem}$ as

\begin{align}
I_{mem} & = 2^{-n} \sum_{a_i, a_i', \{a'_{j\neq i}\}} W_{a_i', \{a'_j\}\to a_i} \ln \frac{\sum_{\{a'_{j \neq i}\}} W_{a_i', \{a'_j\}\to a_i}}{2^{-1} \sum_{a_i', \{a'_{j\neq i}\}} W_{a_i', \{a'_j\}\to a_i}} \notag \\
& = 2^{-n} \sum_{\{a'_{j\neq i}\}} W_{0, \{a'_j\}\to 0} \ln \frac{1/2 - \lambda_0}{2^{-1} \sum_{a_i', \{a'_{j\neq i}\}} W_{a_i', \{a'_j\}\to 0}} \notag \\
& + 2^{-n} \sum_{\{a'_{j\neq i}\}} W_{0, \{a'_j\}\to 1} \ln \frac{\lambda_0}{2^{-1} \sum_{a_i', \{a'_{j\neq i}\}} W_{a_i', \{a'_j\}\to 1}} \notag \\
& + 2^{-n} \sum_{\{a'_{j\neq i}\}} W_{1, \{a'_j\}\to 0} \ln \frac{1/2 - \lambda_1}{2^{-1} \sum_{a_i', \{a'_{j\neq i}\}} W_{a_i', \{a'_j\}\to 0}} \notag \\
& + 2^{-n} \sum_{\{a'_{j\neq i}\}} W_{1, \{a'_j\}\to 1} \ln \frac{\lambda_1}{2^{-1} \sum_{a_i', \{a'_{j\neq i}\}} W_{a_i', \{a'_j\}\to 1}} \notag \\
& = (1/2 - \lambda_0) \ln \frac{1 - 2\lambda_0}{1-\lambda} + \lambda_0 \ln \frac{2\lambda_0}{\lambda} + (1/2 - \lambda_1) \ln \frac{1 - 2\lambda_1}{1-\lambda} + \lambda_1 \ln \frac{2 \lambda_1}{\lambda}
\end{align}

The third and last piece of information we consider is \emph{transfer}, defined as

\begin{equation}
I_{trans} := I(a_i, t; a_k, t-1) .
\end{equation}
where $a_j$ is any neighbor of $a_i$ different from $a_i$ itself. Its computation is quite similar to that of memory and we get

\begin{align}
I_{trans} & = \sum_{a_i, a'_k} P(a_i, t;a'_k, t-1) \ln \frac{P(a_i, t;a'_k, t-1)}{P(a_i, t)P(a'_k, t-1)} \notag \\
& = \sum_{a_i, \{a'_j\}} P(a_i, t;\{a'_j\}, t-1) \ln \frac{\sum_{\{a'_{j \neq k}\}} P(a_i, t;\{a'_j\}, t-1)}{P(a_i, t)P(a'_k, t-1)} \notag \\
& = 2^{-n} \sum_{a_i, \{a'_j\}} W_{\{a_j'\} \to a_i} \ln \frac{\sum_{\{a'_{j \neq k}\}} W_{\{a_j'\} \to a_i}}{2^{-1} \sum_{\{a_j'\}} W_{\{a_j'\} \to a_i}}
\end{align}
We can follow the same way as previously by defining a set of parameters $(\lambda_0^{(L)}, \lambda_1^{(L)}, \lambda_0^{(R)}, \lambda_1^{(R)})$ where the superscript index $L$ or $R$ refers to the left of right neighbor respectively. In other words, $\lambda_0^{(L)}$ is the fraction of ones in the lookup table when the left neighbor is zero, \emph{etc}. As for memory we obviously have $\lambda_0^{(L)} + \lambda_1^{(L)} = \lambda$ and $\lambda_0^{(R)} + \lambda_1^{(R)} = \lambda$. The reasoning exposed for the memory information may be carried over without modification so as to get finally

\begin{align}
I_{trans}^{(L)} & = (1/2 - \lambda_0^{(L)}) \ln \frac{1 - 2\lambda_0^{(L)}}{1-\lambda} + \lambda_0^{(L)} \ln \frac{2\lambda_0^{(L)}}{\lambda} \notag \\
& \qquad + (1/2 - \lambda_1^{(L)}) \ln \frac{1 - 2\lambda_1^{(L)}}{1-\lambda} + \lambda_1^{(L)} \ln \frac{2 \lambda_1^{(L)}}{\lambda}
\end{align}

\begin{align}
I_{trans}^{(R)} & = (1/2 - \lambda_0^{(R)}) \ln \frac{1 - 2\lambda_0^{(R)}}{1-\lambda} + \lambda_0^{(R)} \ln \frac{2\lambda_0^{(R)}}{\lambda} \notag \\
& \qquad + (1/2 - \lambda_1^{(R)}) \ln \frac{1 - 2\lambda_1^{(R)}}{1-\lambda} + \lambda_1^{(R)} \ln \frac{2 \lambda_1^{(R)}}{\lambda}
\end{align}

\paragraph{The case of Rule 110}

As an example consider Rule 110, which is an elementary cellular automaton which is well known for its complex behavior. The lookup table of this dynamics is given by

\begin{equation}
W_{a_{i-1}'a_i'a_{i+1}'\to a_i}=\begin{array}{c|cc|}
    & 0 & 1\\ \hline
000 & 1 & 0 \\
001 & 0 & 1\\
010 & 0 & 1\\
011 & 0 & 1\\
100 & 1 & 0\\
101 & 0 & 1\\
110 & 0 & 1\\
111 & 1 & 0\\
\end{array}
\end{equation}
One immediately gets $\lambda=5/8$, $\lambda_0=\lambda_1^{(L)}=\lambda_0^{(R)}=1/4$ and $\lambda_1=\lambda_0^{(L)}=\lambda_1^{(R)}=3/8$. It results that

\begin{equation}
I_{tot}^{110} = - \frac{5}{8} \ln \frac{5}{8} - \frac{3}{8} \ln \frac{3}{8} \approx 0.66156
\end{equation}

\begin{align}
I_{mem} & = (1/2 - \lambda_0) \ln \frac{1 - 2\lambda_0}{1-\lambda} + \lambda_0 \ln \frac{2\lambda_0}{\lambda} + (1/2 - \lambda_1) \ln \frac{1 - 2\lambda_1}{1-\lambda} + \lambda_1 \ln \frac{2 \lambda_1}{\lambda} \notag \\
& = \frac{1}{4} \ln \frac{4}{3} + \frac{1}{4} \ln \frac{4}{5} + \frac{1}{8} \ln \frac{2}{3} + \frac{3}{8} \ln \frac{6}{5} \approx 0.03382
\end{align}

\begin{align}
I_{trans}^{(L)} & = (1/2 - \lambda_0^{(L)}) \ln \frac{1 - 2\lambda_0^{(L)}}{1-\lambda} + \lambda_0^{(L)} \ln \frac{2\lambda_0^{(L)}}{\lambda} \notag \\
& \qquad + (1/2 - \lambda_1^{(L)}) \ln \frac{1 - 2\lambda_1^{(L)}}{1-\lambda} + \lambda_1^{(L)} \ln \frac{2 \lambda_1^{(L)}}{\lambda} \notag \\
& = \frac{1}{8} \ln \frac{2}{3} + \frac{3}{8} \ln \frac{6}{5} + \frac{1}{4} \ln \frac{4}{3} + \frac{1}{4} \ln \frac{4}{5} \approx 0.03382
\end{align}

\begin{align}
I_{trans}^{(R)} & = (1/2 - \lambda_0^{(R)}) \ln \frac{1 - 2\lambda_0^{(R)}}{1-\lambda} + \lambda_0^{(R)} \ln \frac{2\lambda_0^{(R)}}{\lambda} \notag \\
& \qquad + (1/2 - \lambda_1^{(R)}) \ln \frac{1 - 2\lambda_1^{(R)}}{1-\lambda} + \lambda_1^{(R)} \ln \frac{2 \lambda_1^{(R)}}{\lambda} \notag \\
& = \frac{1}{4} \ln \frac{4}{3} + \frac{1}{4} \ln \frac{4}{5} + \frac{1}{8} \ln \frac{2}{3} + \frac{3}{8} \ln \frac{6}{5} \approx 0.03382
\end{align}

\paragraph{Discussion about the Langton parameter relation}

\begin{figure}
\includegraphics[scale=1]{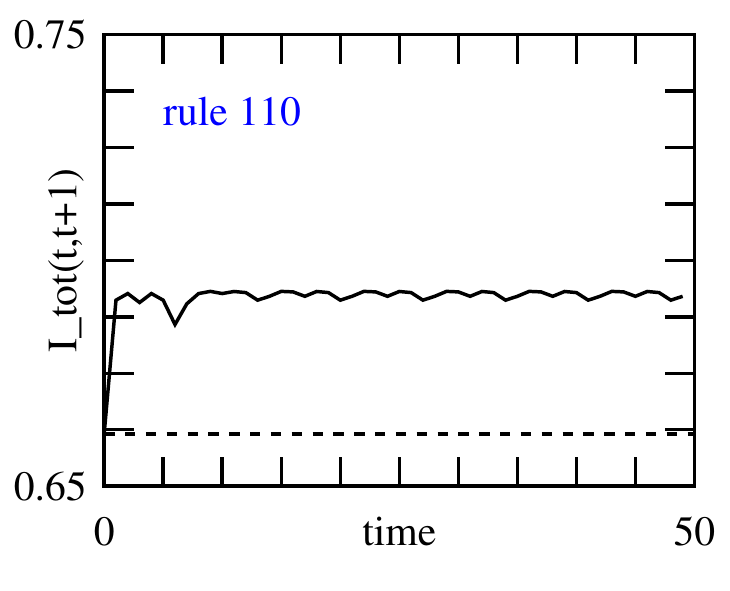}
\caption{\small Total information evaluated for rule 110 over 50 time steps, starting from an uncorrelated initial state.}
\label{Total-Info-R110}
\end{figure}

\begin{figure}
\includegraphics[scale=1]{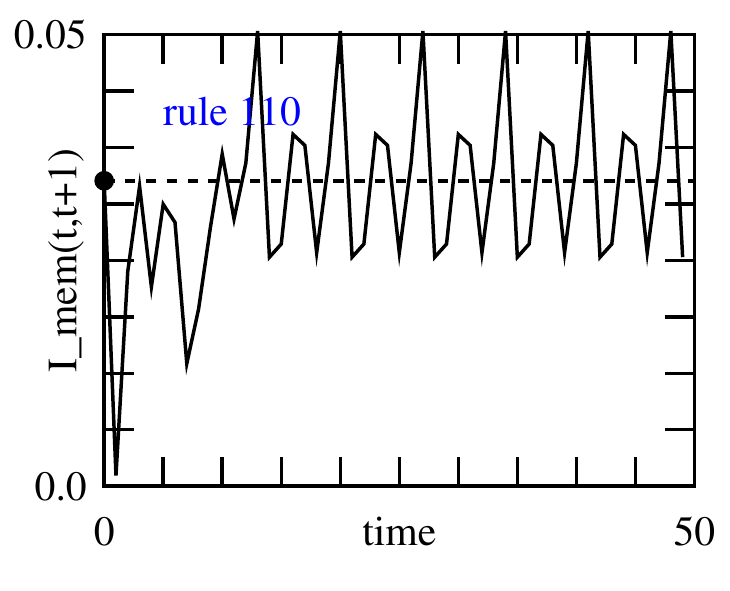}
\caption{\small Memory information evaluated for rule 110 over 50 time steps, starting from an uncorrelated initial state.}
\label{Memory-R110}
\end{figure}

\begin{figure}
\includegraphics[scale=1]{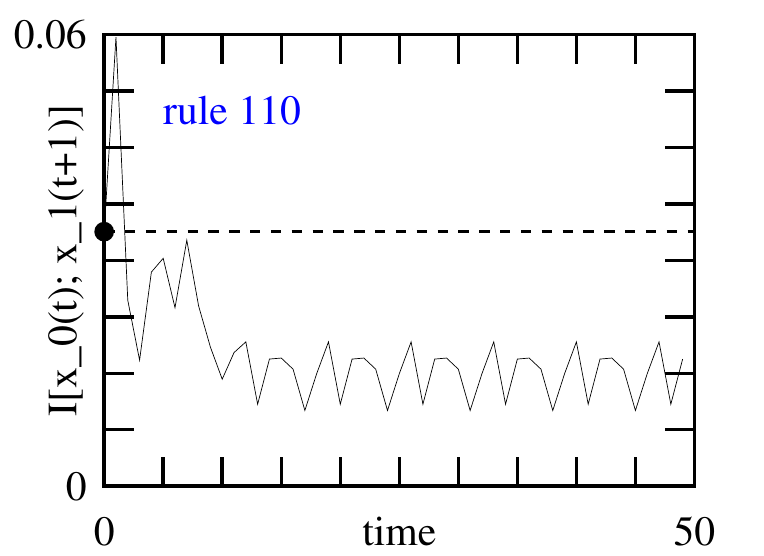}
\caption{\small Left-transfer information evaluated for rule 110 over 50 time steps, starting from an uncorrelated initial state.}
\label{TransferLeft-R110}
\end{figure}

\begin{figure}
\includegraphics[scale=1]{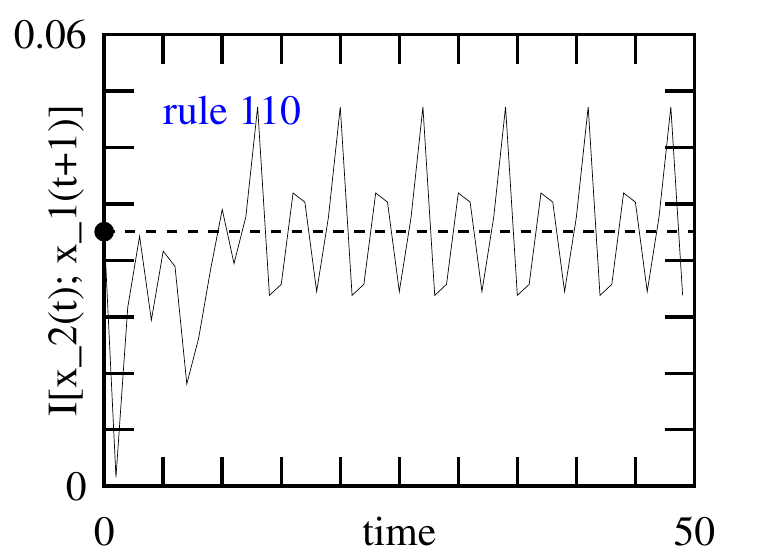}
\caption{\small Right-transfer information evaluated for rule 110 over 50 time steps, starting from an uncorrelated initial state.}
\label{TransferRight-R110}
\end{figure}

So far we have proved that the basic quantities of information processing could be expressed in terms of generalized Langton parameters. This is nevertheless true only when in the initial state the nodes are completely uncorrelated. When the nodes are initially correlated, the values are significantly shifted. Figs.~\ref{Total-Info-R110}, \ref{Memory-R110}, \ref{TransferLeft-R110} and \ref{TransferRight-R110} show the behavior of the total, memory and left- and right-transfer information respectively after the nodes have become correlated. The dotted line pictures the value of each information quantity in the uncorrelated state, highlighting the shift between this value and the corresponding value in the stationary regime.

\subsection*{Robustness of financial cube plots}

\begin{figure}
\includegraphics[width=1.0\textwidth]{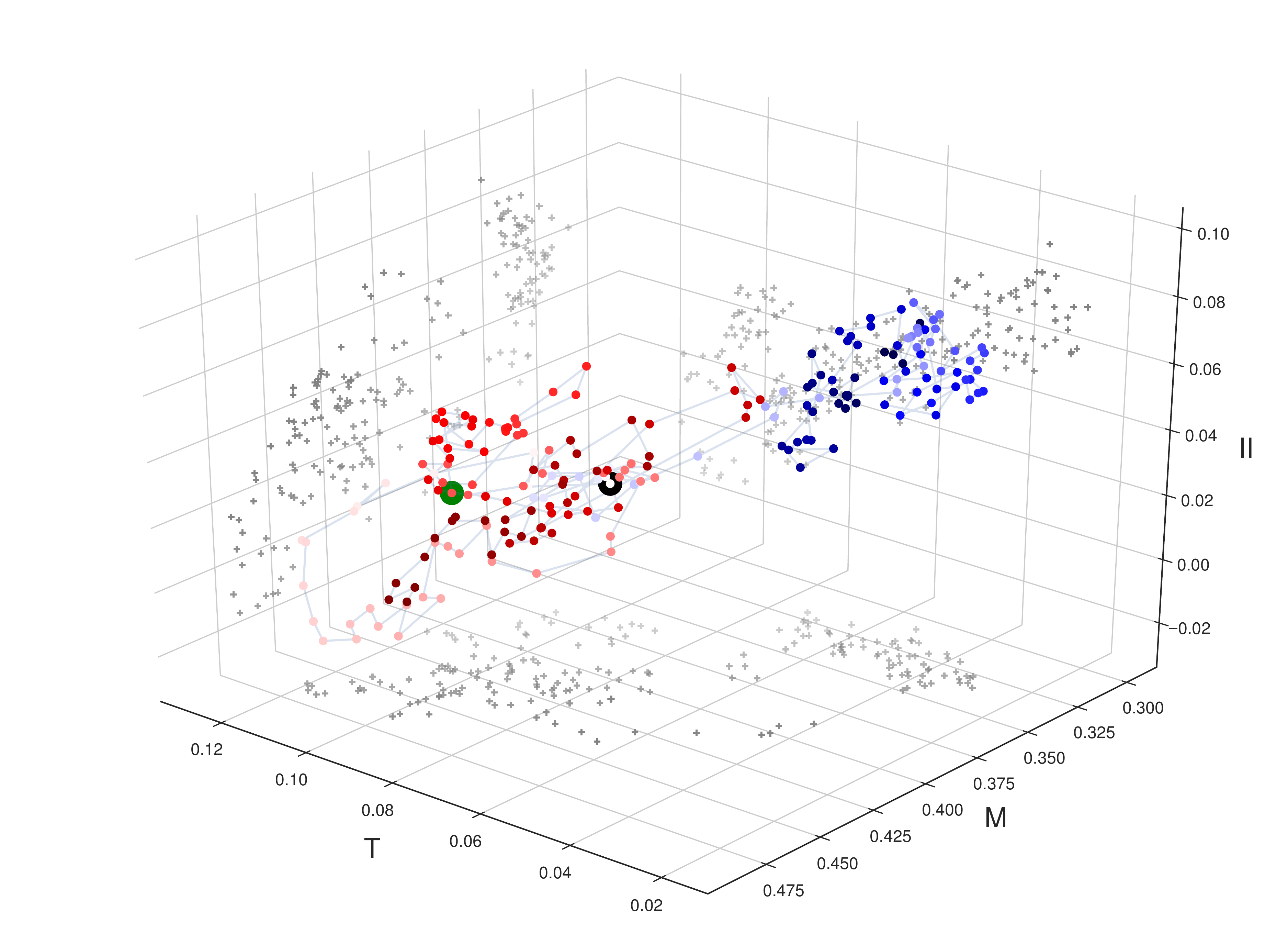}
\caption{200 time points showing the progression of the three information features memory (M), transfer (T), and integration (II) computed with a time delay of 1 day (similar to $t=1$ for ECA). The color indicates the time difference with September 15, 2008 (big black dot), which we consider the starting point of the 2008 crisis, from dark blue (long before) to dark red (long after) and white at the crisis date. The data spans from 1999-01-01 through 2017-04-21; the large green dot is the last time point also present in the IRS data in 2011. In this information space we clearly observe signs of a two attractor regimes separated by a sudden regime shift. The shift is preceded by a slow directed, non-stationary progression in the pre-crisis (blue) regime, resembling the dynamics of a tipping point~\cite{scheffer_early-warning_2009}. Mutual information is calculated using a sliding window of $w=800$ days.}
\label{fig:3dscatter_fx800}
\end{figure}

\begin{figure}
\centering
\includegraphics[width=0.9\textwidth]{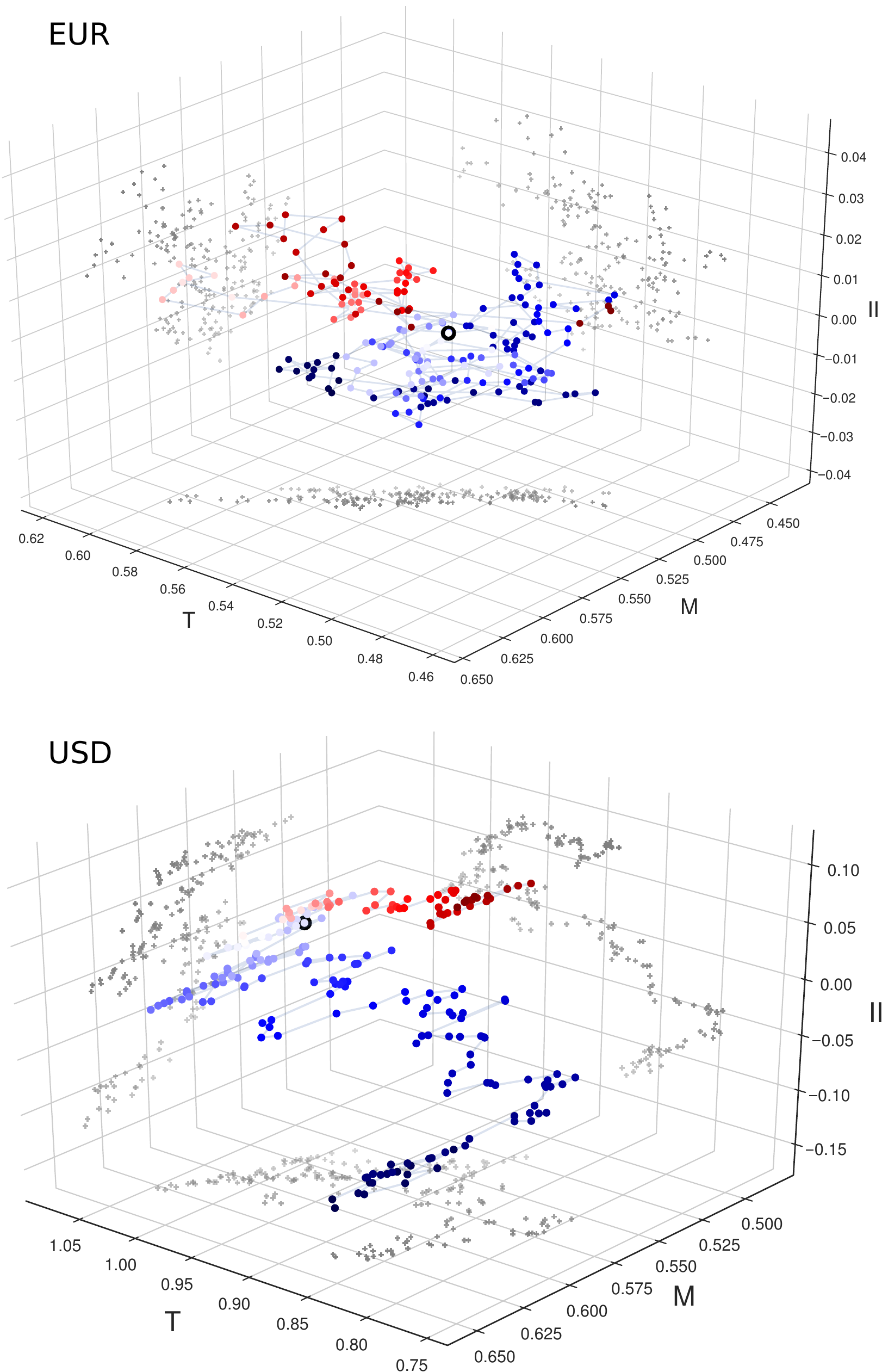}

\caption{200 time points showing the progression of the three information features memory (M), transfer (T), and integration (II) computed with a time delay of 1 day (similar to $t=1$ for ECA). The color indicates the time difference with September 15, 2008 (big black dot), which we consider the starting point of the 2008 crisis, from dark blue (long before) to dark red (long after) and white at the crisis date. The data spans more than twelve years: the EUR data from 1998-01-12 to 2011-08-12 and the USD data from 1999-29-04 to 2011-06-06. In the EUR market two regimes can be identified which are well-separated but close together; in the USD market we see instead a steady, prolonged progression in a similar direction. Mutual information is calculated using a sliding window of $w=800$ days.}
\label{fig:3dscatter_irs800}
\end{figure}

\begin{figure}
\includegraphics[width=1.0\textwidth]{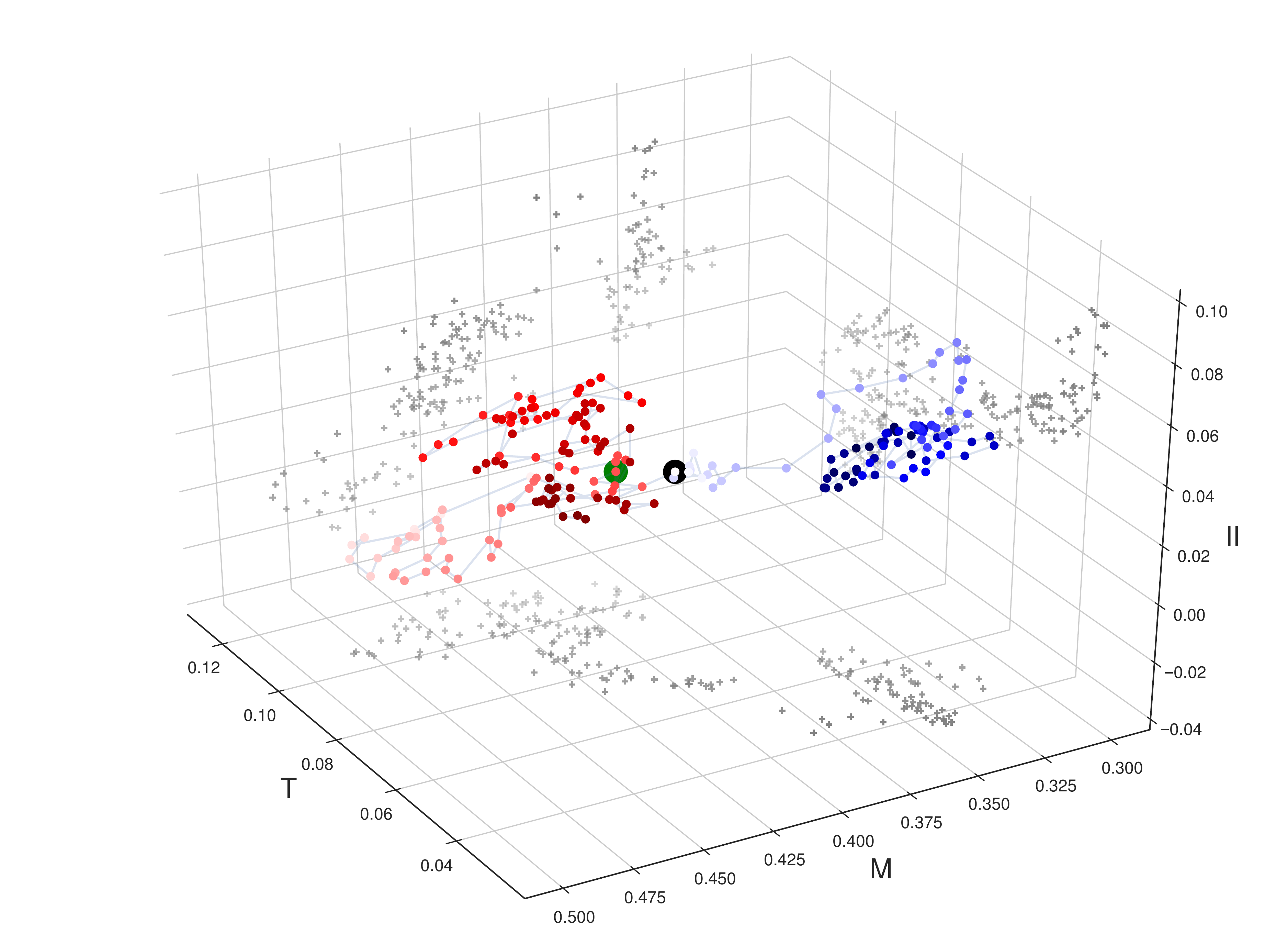}
\caption{200 time points showing the progression of the three information features memory (M), transfer (T), and integration (II) computed with a time delay of 1 day (similar to $t=1$ for ECA). The color indicates the time difference with September 15, 2008 (big black dot), which we consider the starting point of the 2008 crisis, from dark blue (long before) to dark red (long after) and white at the crisis date. The data spans from 1999-01-01 through 2017-04-21; the large green dot is the last time point also present in the IRS data in 2011. In this information space we clearly observe signs of a two attractor regimes separated by a sudden regime shift. The shift is preceded by a slow directed, non-stationary progression in the pre-crisis (blue) regime, resembling the dynamics of a tipping point~\cite{scheffer_early-warning_2009}. Mutual information is calculated using a sliding window of $w=1000$ days.}
\label{fig:3dscatter_fx1000}
\end{figure}

\begin{figure}[htp]

\centering

%
\includegraphics[width=0.9\linewidth]{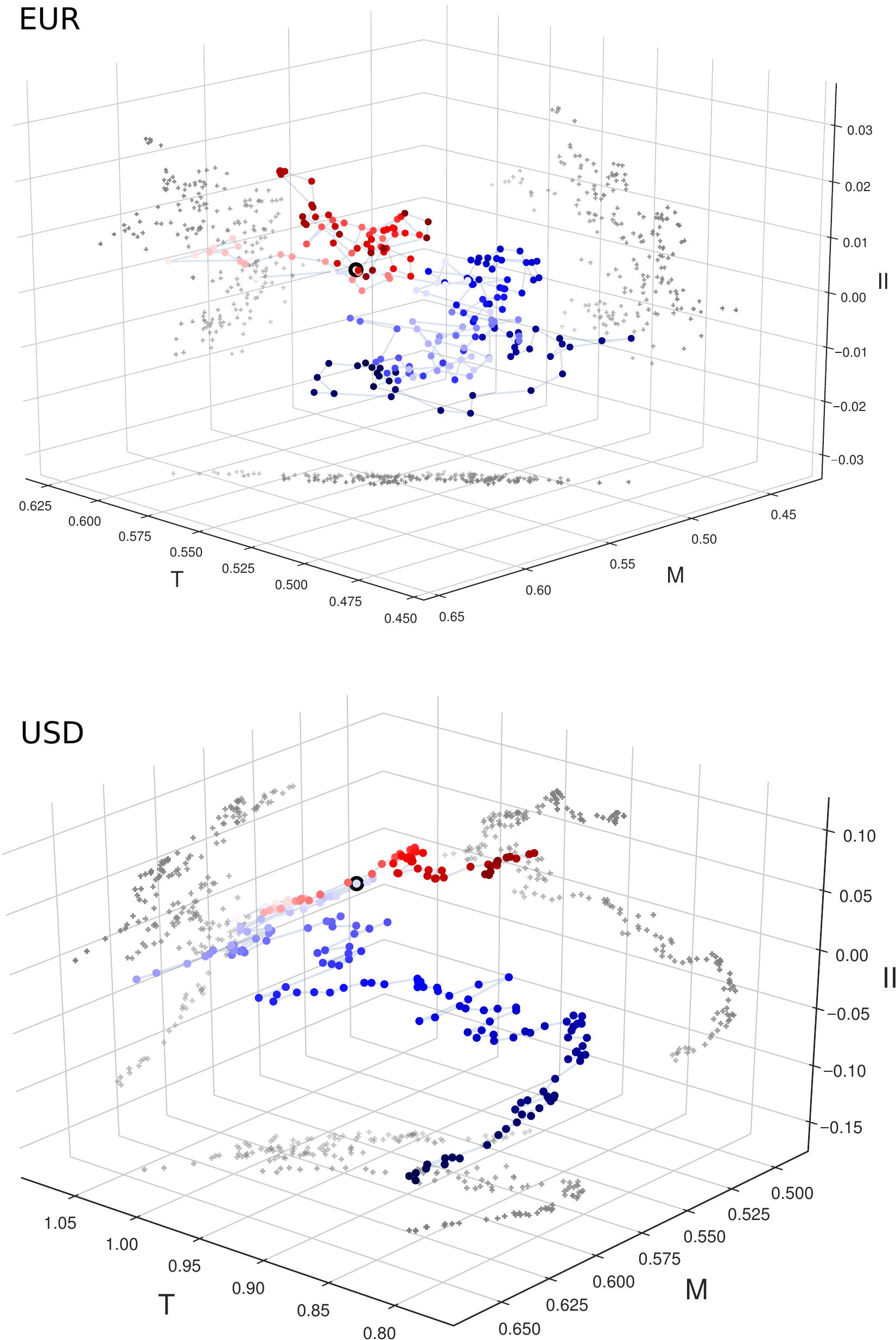}

\caption{200 time points showing the progression of the three information features memory (M), transfer (T), and integration (II) computed with a time delay of 1 day (similar to $t=1$ for ECA). The color indicates the time difference with September 15, 2008 (big black dot), which we consider the starting point of the 2008 crisis, from dark blue (long before) to dark red (long after) and white at the crisis date. The data spans more than twelve years: the EUR data from 1998-01-12 to 2011-08-12 and the USD data from 1999-29-04 to 2011-06-06. In the EUR market two regimes can be identified which are well-separated but close together; in the USD market we see instead a steady, prolonged progression in a similar direction. Mutual information is calculated using a sliding window of $w=1000$ days.}

\label{fig:3dscatter_irs1000}

\end{figure}

\paragraph{Sliding window size.} Here we show result plots for the financial time series (Fig.~\ref{fig:3dscatter_fx} and Fig.~\ref{fig:3dscatter_irs} in the main text) for lower sliding window size ($w=1400$ data points in main text) in order to demonstrate that the results in the main text are not obtained by carefully fine-tuning this parameter. The lower the sliding window size, the fewer datapoints and thus the more difficult it is to accurately estimate mutual information. We show here $w=800$ and $w=1000$ in Figs.~\ref{fig:3dscatter_fx800} through \ref{fig:3dscatter_irs1000} to illustrate how the main text's results already start to appear at these significantly lower values.

\section*{Acknowledgments}
PMAS and RQ acknowledge the financial support of the Future and Emerging Technologies (FET) program within Seventh Framework Programme (FP7) for Research of the European Commission, under the FET- Proactive grant agreement TOPDRIM, number FP7-ICT-318121. PMAS, RQ, and OHS also acknowledge the financial support of the Future and Emerging Technologies (FET) program within Seventh Framework Programme (FP7) for Research of the European Commission, under the FET-Proactive grant agreement Sophocles, number FP7-ICT-317534. PMAS acknowledges the support of the Russian Scientific Foundation, Project number 14-21-00137.

\section*{Conflicts of interests}
The authors declare that there is no conflict of interest regarding the publication of this paper.


%
%
%
%
%

\bibliographystyle{plos2015}

\bibliography{references}

\begin{thebibliography}{10}

\bibitem{cover_elements_1991}
Cover TM, Thomas JA.
\newblock Elements of information theory. vol.~6.
\newblock Wiley-Interscience; 1991.

\bibitem{lizier_information_2010}
Lizier JT, Prokopenko M, Zomaya AY.
\newblock Information modification and particle collisions in distributed
  computation.
\newblock Chaos: An Interdisciplinary Journal of Nonlinear Science.
  2010;20(3):037109.

\bibitem{lizier_information_2008}
Lizier JT, Prokopenko M, Zomaya AY.
\newblock The {Information} {Dynamics} of {Phase} {Transitions} in {Random}
  {Boolean} {Networks}.
\newblock In: {ALIFE}; 2008. p. 374--381.

\bibitem{beer_information_2014}
Beer RD, Williams PL.
\newblock Information processing and dynamics in minimally cognitive agents.
\newblock Cognitive science. 2014;.

\bibitem{izquierdo_information_2015}
Izquierdo EJ, Williams PL, Beer RD.
\newblock Information flow through a model of the {C}. elegans klinotaxis
  circuit.
\newblock arXiv preprint arXiv:150204262. 2015;.

\bibitem{bar-yam_computationally_2013}
Bar-Yam Y, Harmon D, Bar-Yam Y.
\newblock Computationally tractable pairwise complexity profile.
\newblock Complexity. 2013;18(5):20--27.

\bibitem{allen_information-theoretic_2014}
Allen B, Stacey BC, Bar-Yam Y.
\newblock An {Information}-{Theoretic} {Formalism} for {Multiscale} {Structure}
  in {Complex} {Systems}.
\newblock arXiv preprint arXiv:14094708. 2014;.

\bibitem{quax_diminishing_2013}
Quax R, Apolloni A, Sloot PMA.
\newblock The diminishing role of hubs in dynamical processes on complex
  networks.
\newblock Journal of The Royal Society Interface. 2013;10(88):20130568.

\bibitem{quax_information_2013-1}
Quax R, Kandhai D, Sloot PMA.
\newblock Information dissipation as an early-warning signal for the {Lehman}
  {Brothers} collapse in financial time series.
\newblock Scientific reports. 2013;3.

\bibitem{kristian_lindgren_information-theoretic_2015}
{Kristian Lindgren}.
\newblock An information-theoretic perspective on coarse-graining, including
  the transition from micro to macro.
\newblock Entropy. 2015;.

\bibitem{james_anatomy_2011}
James RG, Ellison CJ, Crutchfield JP.
\newblock Anatomy of a bit: {Information} in a time series observation.
\newblock Chaos: An Interdisciplinary Journal of Nonlinear Science.
  2011;21(3):037109.

\bibitem{williams_nonnegative_2010}
Williams PL, Beer RD.
\newblock Nonnegative decomposition of multivariate information.
\newblock arXiv preprint arXiv:10042515. 2010;.

\bibitem{olbrich_information_2015}
Olbrich E, Bertschinger N, Rauh J.
\newblock Information {Decomposition} and {Synergy}.
\newblock Entropy. 2015;17(5):3501.
\newblock Available from: \url{http://www.mdpi.com/1099-4300/17/5/3501}.

\bibitem{quax_synergy}
Quax R, Har-Shemesh O, Sloot PMA.
\newblock Quantifying Synergistic Information Using Intermediate Stochastic
  Variables.
\newblock Entropy. 2017;19(2).

\bibitem{chliamovitch2014assessing}
Chliamovitch G, Chopard B, Velasquez L.
\newblock Assessing complexity by means of maximum entropy models.
\newblock arXiv preprint arXiv:14080368. 2014;.

\bibitem{griffith_intersection_2014}
Griffith V, Chong EKP, James RG, Ellison CJ, Crutchfield JP.
\newblock Intersection {Information} {Based} on {Common} {Randomness}.
\newblock Entropy. 2014;16(4):1985--2000.
\newblock Available from: \url{http://www.mdpi.com/1099-4300/16/4/1985}.

\bibitem{griffith_synergy_entropy_2015}
Griffith V, Ho T.
\newblock Quantifying Redundant Information in Predicting a Target Random
  Variable.
\newblock Entropy. 2015;17(7):4644.
\newblock Available from: \url{http://www.mdpi.com/1099-4300/17/7/4644}.

\bibitem{wolfram_new_2002}
Wolfram S.
\newblock A {New} {Kind} of {Science}.
\newblock Wolfram Media; 2002.
\newblock Available from:
  \url{http://www.amazon.com/exec/obidos/ASIN/1579550088/ref=nosim/rds-20}.

\bibitem{wolframalpha}
Wolfram|Alpha: Computational Knowledge Engine;.
\newblock September 7, 2015.
\newblock \url{http://www.wolframalpha.com/}.

\bibitem{battiti1994using}
Battiti R.
\newblock Using mutual information for selecting features in supervised neural
  net learning.
\newblock Neural Networks, IEEE Transactions on. 1994;5(4):537--550.

\bibitem{chow2005estimating}
Chow TW, Huang D.
\newblock Estimating optimal feature subsets using efficient estimation of
  high-dimensional mutual information.
\newblock Neural Networks, IEEE Transactions on. 2005;16(1):213--224.

\bibitem{lizier_towards_2013}
Lizier JT, Flecker B, Williams PL.
\newblock Towards a synergy-based approach to measuring information
  modification.
\newblock In: Artificial {Life} ({ALIFE}), 2013 {IEEE} {Symposium} on. IEEE;
  2013. p. 43--51.

\bibitem{beer2015information}
Beer RD, Williams PL.
\newblock Information processing and dynamics in minimally cognitive agents.
\newblock Cognitive science. 2015;39(1):1--38.

\bibitem{schneidman_synergy_2003}
Schneidman E, Bialek W, Berry MJ.
\newblock Synergy, redundancy, and independence in population codes.
\newblock the Journal of Neuroscience. 2003;23(37):11539--11553.

\bibitem{newman_structure_2003}
Newman MEJ.
\newblock The {Structure} and {Function} of {Complex} {Networks}.
\newblock SIAM Rev Soc Ind Appl Math. 2003 Jan;45:167--256.

\bibitem{kraskov_estimating_2004}
Kraskov A, Stögbauer H, Grassberger P.
\newblock Estimating mutual information.
\newblock Physical review E. 2004;69(6):066138.

\bibitem{quax2013information}
Quax R, Kandhai D, Sloot PM.
\newblock Information dissipation as an early-warning signal for the Lehman
  Brothers collapse in financial time series.
\newblock Scientific reports. 2013;3:1898.

\bibitem{langton1990computation}
Langton CG.
\newblock Computation at the edge of chaos: phase transitions and emergent
  computation.
\newblock Physica D: Nonlinear Phenomena. 1990;42(1):12--37.

\bibitem{culik1988undecidability}
Culik~II K, Yu S.
\newblock Undecidability of CA classification schemes.
\newblock Complex Systems. 1988;2(2):177--190.

\bibitem{sutner2012classify}
Sutner K.
\newblock Computational classification of cellular automata.
\newblock International Journal of General Systems. 2012;41(6):595--607.
\newblock Available from: \url{http://dx.doi.org/10.1080/03081079.2012.695899}.

\bibitem{scheffer_early-warning_2009}
Scheffer M, Bascompte J, Brock WA, Brovkin V, Carpenter SR, Dakos V, et~al.
\newblock Early-warning signals for critical transitions.
\newblock Nature. 2009;461(7260):53--59.

\bibitem{Melvin20091317}
Melvin M, Taylor MP.
\newblock The crisis in the foreign exchange market.
\newblock Journal of International Money and Finance. 2009;28(8):1317 -- 1330.
\newblock The Global Financial Crisis: Causes, Threats and Opportunities.
\newblock Available from:
  \url{http://www.sciencedirect.com/science/article/pii/S0261560609000953}.

\bibitem{saunders2010credit}
Saunders A, Allen L.
\newblock Credit risk management in and out of the financial crisis: new
  approaches to value at risk and other paradigms. vol. 528.
\newblock John Wiley \& Sons; 2010.

\bibitem{graaf2014cva}
De~Graaf CS, Feng Q, Kandhai D, Oosterlee CW.
\newblock Efficient computation of exposure profiles for counterparty credit
  risk.
\newblock International Journal of Theoretical and Applied Finance.
  2014;17(04):1450024.

\bibitem{lizier_local_2012}
Lizier JT.
\newblock The local information dynamics of distributed computation in complex
  systems.
\newblock Springer; 2012.

\end{thebibliography}

\end{document}